\documentclass[prb, reprint, twocolumn, superscriptaddress, amsfonts, amssymb, amsmath]{revtex4-2}
\usepackage[english]{babel}
\usepackage{hyperref}
\usepackage{graphicx}
\usepackage{physics}
\usepackage{bm}

\begin{document}


\title{Topological signatures in a weakly dissipative Kitaev chain of finite length}

\date{\today}

\author{Antonio \surname{D'Abbruzzo}}
\affiliation{Dipartimento di Fisica dell'Università di Pisa, Largo Pontecorvo 3, I-56127 Pisa, Italy}

\author{Davide \surname{Rossini}}
\affiliation{Dipartimento di Fisica dell'Università di Pisa, Largo Pontecorvo 3, I-56127 Pisa, Italy}
\affiliation{INFN, Sezione di Pisa, Largo Pontecorvo 3, I-56127 Pisa, Italy}

\begin{abstract}
  We construct a global Lindblad master equation for a Kitaev quantum wire of finite length,
  weakly coupled to an arbitrary number of thermal baths, within the Born-Markov and secular approximations.
  We find that the coupling of an external bath to more than one lattice site generates
  quantum interference effects, arising from the possibility of fermions to tunnel through multiple paths.
  In the presence of two baths at different temperatures and/or chemical potentials,
  the steady-state particle current can be expressed through the Landauer-B\"uttiker formula, as in
  a ballistic transport setup, with the addition of an anomaly factor associated with the presence
  of the $p$-wave pairing in the Kitaev Hamiltonian.
  Such a factor is affected by the ground-state properties of the chain,
  being related to the finite-size equivalent of its Pfaffian topological invariant.
\end{abstract}

\maketitle


\section{Introduction}

Characterizing the nonequilibrium features of quantum many-body systems is one
of the cornerstone problems in modern condensed matter physics.
This task becomes even more complex in the presence of interactions with an external
environment, where even the system-bath modelization presents intrinsic conceptual issues.

Open quantum systems are generally described by density operators, whose
time evolution is governed by suitable master equations~\cite{Petruccione-07, Alicki-07}.
The most studied setting lies within the weak coupling hypothesis, leading to a Markovian
behavior that is captured by the well-known Lindblad equation~\cite{Lindblad-76}.
Relaxing the above condition may inherit memory effects that give rise to more complex situations
governed by integro-differential master equations, whose description poses some subtleties
which are still object of research.
Even restricting to Markovian master equations, a proper description of interactions with
the environment generally requires the full knowledge of the eigendecomposition of the system Hamiltonian,
which is feasible only for special classes of many-body systems or in specific
situations~\cite{Harbola-06, Santos-16, Benatti-20, Cattaneo-20, Dorn-21, DAbbruzzo-21}.
In fact, most studies in the many-body realm stick with local forms of Lindblad jump
operators in the master equations (see, e.g., Ref.~\cite{Sieberer-16} and references therein).
Even if this hypothesis can be considered safe in several important situations,
as in quantum optical implementations~\cite{Muller-12},
this is not true for many other applications in condensed matter physics, as in solid state devices.

Here we discuss the role that an external thermal environment may have on a superconducting nanowire,
under the regime of weak and Markovian coupling. Namely, we study the quantum dynamics
of the Kitaev chain governed by a global Lindblad master equation,
which can be microscopically derived in a self-consistent way
using the spectral decomposition obtained through a Bogoliubov approach~\cite{DAbbruzzo-21}.
This treatment fully captures the effects induced on the system by the presence of
weak interactions with an arbitrary number of thermal baths.

The Kitaev model is a simple toy model of a parity-conserving quadratic fermionic chain~\cite{Kitaev-01}.
Nonetheless, it constitutes a prototypical example to describe one-dimensional
spin-polarized superconductors, with the appearance of a nontrivial topology related
to the emergence of edge states, in the form of an unbound pair
of Majorana fermions~\cite{DasSarma-10, vonOppen-10}.
For this reason, it has been recently object of intense theoretical and experimental
investigations~\cite{Alicea, Franz}, while the debate on the detection of Majorana modes, e.g.,
in the form of a zero-bias peak in the tunneling conductance is still vivid~\cite{Mourik-12, Zhang-19}.
Understanding the topological features of matter through this kind of model is also useful to develop
new platforms for quantum computation and information processing~\cite{Nayak-08}.
We finally recall that, by tuning the various Hamiltonian parameters, one can study
the critical properties of the emerging topological quantum phase transition (QPT),
lying in the two-dimensional Ising universality class~\cite{RV-21}.

Due to the broad significance of the Kitaev chain in several contexts, a proper comprehension of
the decoherence effects emerging when immersed into some environment is relevant
not only from an abstract point of view, as in topology physics or in critical phenomena,
but also for practical purposes in quantum simulation of Majorana nanowires~\cite{DasSarma-17}.
In this respect, previous investigations shed light on the role of dissipative and dephasing noise,
which induce certain types of transitions between the eigenstates of the system~\cite{Ng-15},
as well as on the dynamical behavior of the system (with emphasis on its topological features)
under time periodic modulation of its Hamiltonian parameters~\cite{Molignini-17},
or in the presence of environments modeled by sets of harmonic oscillators~\cite{Huang-19} or
different types of local Lindblad jump operators~\cite{vanCaspel-19, Dutta-20, Cooper-20, Cooper-21}.
Other works focused on the emergence of topological phases in the presence
of Lindbladian dissipation using quantum-information based alternative approaches~\cite{Delgado-12, Delgado-14}.
Here we are interested in providing simple and rigorous results in the global (i.e., nonlocal)
Markovian scenario.

We show that the dynamics predicted by our model manifests quantum interference phenomena,
originating when a thermal environment is coupled to the superconducting nanowire through more
than one lattice site contacts, due to the creation of multiple paths where a fermion can tunnel.
We also address quantum transport in a minimal setting, where
the Kitaev chain is coupled to two baths at different temperatures and/or chemical potentials.
The fermionic particle current establishing in the nanowire can be described through an expression
which closely resembles the structure of the Landauer-B{\"u}ttiker scattering-matrix formula
for ballistic transport, except for the presence of a multiplicative factor which we call
the anomaly factor.
In fact, the latter originates from the anomalous terms of the system Hamiltonian,
bringing to important qualitative difference with respect to the case of ballistic channels:
in the case of the Kitaev chain, the anomaly factor is able to keep memory of the
phase of the system, a fact which enables to relate it to bulk-edge correspondence arguments.

The paper is organized as follows. In Sec.~\ref{sec:Model} we present the model
under investigation, starting from the Hamiltonian of the Kitaev chain with open
boundary conditions (Sec.~\ref{sec:kitaev}) and then adding a weak coupling to
thermal baths in the Born-Markov approximation, which can be microscopically treated
through a master equation in the Lindblad form (Sec.~\ref{sec:MasterEq}),
provided a full secular approximation is enforced (Sec.~\ref{sec:secular}).
In Sec.~\ref{sec:OneBath} we discuss the case in which a given number of sites in
the Kitaev chain is coupled to a single common bath, highlighting how the system-environment
coupling enables to generate interference patterns in the fermionic tunneling among the sites.
We then discuss, in Sec.~\ref{sec:TwoBaths}, the case of a chain coupled to two thermal baths.
The anomaly factor appearing in the expression for the particle current depends
on the different phases of the system, thus affecting the transport properties.
Finally, in Sec.~\ref{sec:Conclusions} we draw our conclusions.
In the appendices, we explicitly show how to diagonalize an open-ended
Kitaev chain with a finite length (Appendix~\ref{appendixA}), and discuss in some detail
the regime of validity of the secular approximation (Appendix~\ref{appendixB}).


\section{Description of the model}
\label{sec:Model}

\subsection{Finite Kitaev chain with open boundaries}
\label{sec:kitaev}

We consider a one-dimensional quantum lattice of $~{N > 1}$ sites, described by the Hamiltonian
\begin{equation}\label{eq:general_kitaev}
   H_\mathcal{S} = - \sum_{j=1}^N \left[ \big( t \, a_j^\dagger a_{j+1} +
   \Delta \, a_j^\dagger a_{j+1}^\dagger + \text{h.c.} \big) + \mu \, a_j^\dagger a_j \right] \, ,
\end{equation}
where $t,\Delta,\mu$ are real parameters and $a_j$ ($a_j^\dagger$) is the annihilation (creation)
operator of a fermionic particle on the $j$-th site. We impose open boundary conditions,
meaning that $a_{N+1} = a^\dagger_{N+1} = 0$. The model in Eq.~\eqref{eq:general_kitaev}
was originally proposed by Kitaev to describe a p-wave superconducting chain,
featuring the emergence of Majorana edge modes~\cite{Kitaev-01}.
In the thermodynamic limit $N \rightarrow \infty$, it undergoes a topological quantum phase
transition (QPT) at the critical points $\mu_c = \pm 2t$,
i.e., for $|\mu| < 2|t|$ a zero-energy mode localized at the ends of the chain appears:
this is a manifestation of a different topological structure of the system,
with respect to the one characterizing the trivial phase for $|\mu| > 2|t|$.

Despite the great amount of studies carried out on the Kitaev chain
(see, e.g., Ref.~\cite{Alicea} and references therein), a complete description
of the associated eigenvalue problem for finite $N$
has been achieved only recently~\cite{Kao-14, Zvyagin-15, Hegde-15, Zeng-19, Leumer-20, Leumer-21}.
For our purposes, it is sufficient to observe that there
exist nontrivial values of the parameters $t,\mu,\Delta$ leading to the appearance of
degeneracies in the excitation spectrum~\cite{Leumer-21}.
Therefore, to avoid complications in deriving the master equation in the dissipative
scenario~\cite{DAbbruzzo-21}, we will limit ourselves to the case $t = \Delta$ and $\mu \neq 0$,
which is known to be degeneracy-free, as discussed below.
Specifically, hereafter we fix the energy scale as $t = \Delta = 1$ and specialize
to the following Hamiltonian with a nonzero real parameter $\mu$:
\begin{equation}
  \label{eq:kitaev}
   H_\mathcal{S} = -\sum_{j=1}^{N-1} \left( a_j^\dagger a_{j+1} + a_j^\dagger a_{j+1}^\dagger
   + \text{h.c.} \right) - \mu \sum_{j=1}^N a_j^\dagger a_j \, .
\end{equation}
This model can be also obtained by applying a Jordan-Wigner transformation to the transverse-field
quantum Ising chain with open boundary conditions~\cite{Pfeuty-70}, where $\mu$
is proportional to the magnitude of the transverse field and the corresponding QPT
separates a paramagnetic from a ferromagnetic phase~\cite{Sachdev}.
However, to avoid any complication deriving from the nonlocal character of the Jordan-Wigner transformation,
we prefer to stick with the purely fermionic interpretation of the Kitaev model.

The Hamiltonian~\eqref{eq:kitaev} is quadratic and can be put in diagonal
form by means of a standard Bogoliubov-Valatin (BV) transformation~\cite{Blaizot-Ripka, Xiao-09}
\begin{equation}
  \label{eq:BV_direct}
   a_j = \sum_k \left( A_{jk} b_k + B_{jk} b_k^\dagger \right) \, ,
\end{equation}
where $b_k$ are new fermionic operators. The (real) coefficient matrices $A,B$ can be chosen
following a method originally proposed by Lieb, Schultz, and Mattis (LSM)~\cite{LSM-61},
in such a way to obtain
\begin{equation}
   H_\mathcal{S} = \sum_k \omega_k b_k^\dagger b_k + \text{const.} \,\, ,
\end{equation}
where $\omega_k$ is a nonnegative real number, representative of a quasiparticle excitation energy.
The diagonalization procedure of Eq.~\eqref{eq:kitaev} for a finite chain length $N$ is not
straightforward~\cite{Leumer-20}. Below we only report the most important results,
which turn out to be useful in the following; the interested reader can find
further details in Appendix~\ref{appendixA}, where thorough derivations are provided.

\begin{figure}
   \centering
   \includegraphics[width=\columnwidth]{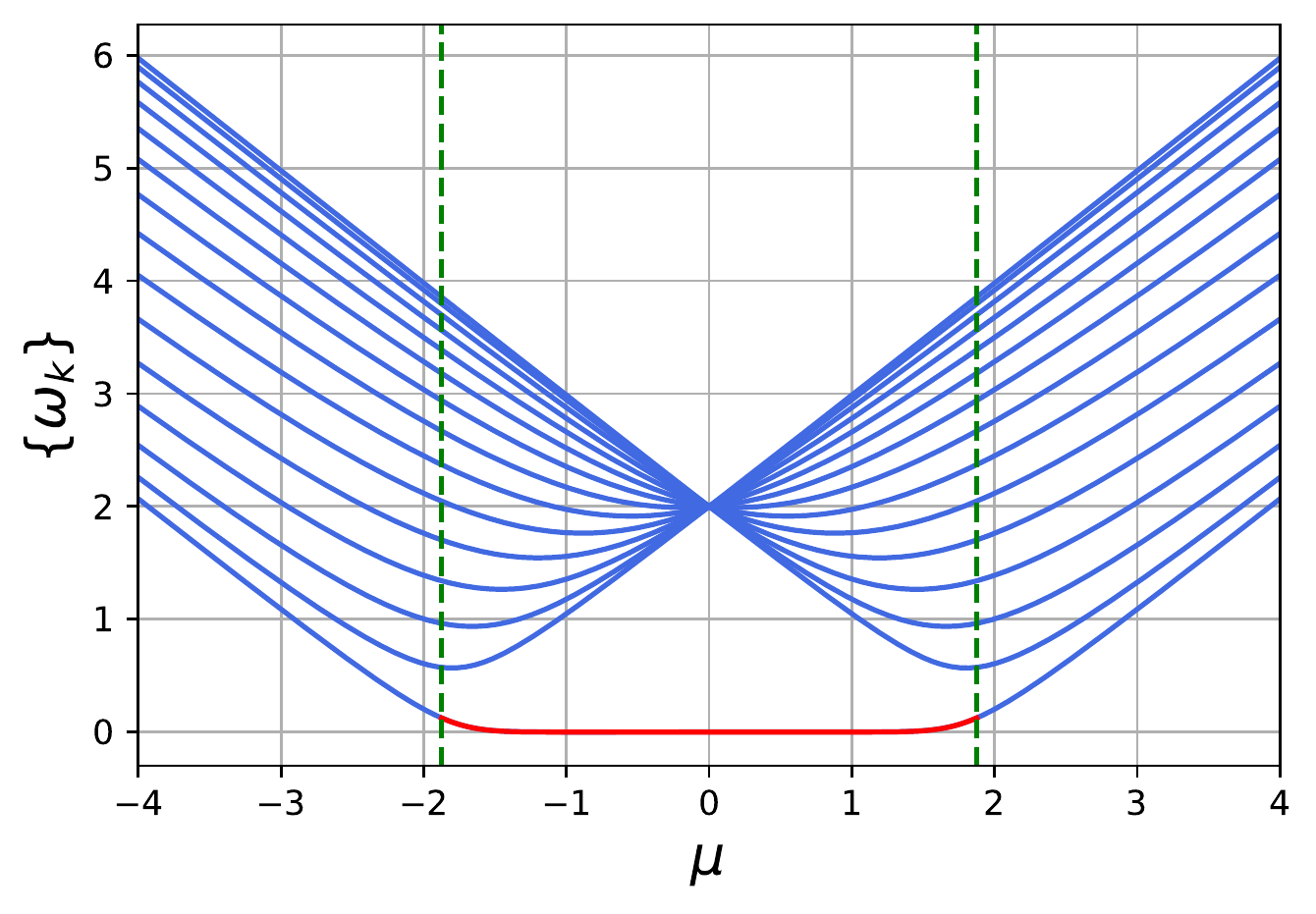}
   \caption{The excitation spectrum~\eqref{eq:spectrum} of the finite-size Kitaev
     Hamiltonian~\eqref{eq:kitaev} as a function of the parameter $\mu$, for $N = 15$.
     Blue curves are associated with bulk modes, while the red curve represents
     the edge mode contribution [see Eq.~\eqref{eq:spectrum_edge}],
     which only appears for $|\mu| < |\mu_{\rm sep}| \equiv 2N/(N+1)$.
     The position of the separation points are indicated with green dashed lines.
     The results have been obtained by diagonalizing the matrix $W$ in Eq.~\eqref{eq:W},
     whose spectrum is $\{\omega_k^2\}$ [see Eq.~\eqref{eq:eigen_problem_W}].}
   \label{fig:spectrum}
\end{figure}

The dispersion relation of the BV quasiparticles reads
\begin{equation}
  \label{eq:spectrum}
   \omega_k^2 = \mu^2 + 4\mu \cos k + 4 \, ,
\end{equation}
where $k$ is, in general, a complex number satisfying the quantization condition
\begin{equation}
  \label{eq:quantization}
   2 \sin kN + \mu \sin \big[ k(N+1) \big] = 0 \, .
\end{equation}
It can be shown that, in order to make $\omega_k^2 \geq 0$, other constraints on $k$ have
to be imposed. In particular, $k$ has to be a real number in the open interval $(0,\pi)$
(we refer to it as a \textit{bulk mode}) or it has the shape $k = \theta(\mu) \pi + i \eta$, where
$\theta(\cdot)$ is the Heaviside step function and $\eta > 0$ (we refer to it as a \textit{edge mode}).
In the latter case, making the identification $k \rightarrow \eta$, one can express the
quasiparticle energy in terms of $\eta$ as
\begin{equation}\label{eq:spectrum_edge}
   \omega_\eta^2 = \mu^2 - 4|\mu| \cosh\eta + 4 \, .
\end{equation}
An analysis of Eq.~\eqref{eq:quantization} shows that there exists a value $|\mu_{\rm sep}| = 2N/(N+1)$
for the chemical potential separating the region of $\mu$ in which the complex-$k$ mode is present
[see Eq.~\eqref{eq:separation_app}]. In the thermodynamic limit $N \rightarrow \infty$, it tends
to the critical point $|\mu_c| = 2$ of the topological QPT, where the complex-$k$ mode is nothing
but the zero-energy Majorana edge mode.
Figure~\ref{fig:spectrum} reports the spectrum $\{\omega_k\}$ as a function of $\mu$ for finite $N$,
showing a clear separation in energy between the bulk modes (blue curves)
and the edge mode (red curve). The data also confirm the previous claim on the nondegenerate
character of the spectrum
(apart from the pathological case $\mu = 0$, which we exclude from the beginning).

For what concerns the coefficients $A_{jk}$, $B_{jk}$ in the BV transformation~\eqref{eq:BV_direct},
defining the so-called LSM matrices $\phi \equiv A + B$ and $\psi \equiv A - B$, the following facts hold.
For real $k$, one finds
\begin{subequations}\label{eq:LSM}
\begin{eqnarray}
   \psi_{jk} & = & C_k \sin (kj) \, , \label{eq:LSM_psi} \\
   \phi_{jk} & = & D_k \sin \big[ k(N+1-j) \big] \, , \label{eq:LSM_phi}
\end{eqnarray}
\end{subequations}
where the normalization constants $C_k$, $D_k$ are real and satisfy the conditions
\begin{subequations}
  \begin{eqnarray}
    C_k^2 & = & D_k^2 \, , \\
   \dfrac{2}{D_k^2} & = & N - \dfrac{\sin (kN)}{\sin k} \cos \big[k(N+1) \big] \, ,
   \label{eq:normalization} \\
   \dfrac{D_k}{C_k} & = & -\dfrac{\mu \sin k}{\omega_k \sin (kN)} \, . \label{eq:normalization2}
\end{eqnarray}
\end{subequations}
These imply $C_k = \pm D_k$, where the relative sign depends on $k, \mu, N$
through Eqs.~\eqref{eq:normalization},\eqref{eq:normalization2}.
In contrast, for the complex-$k$ mode, making the identification $k \rightarrow \eta$, one finds
\begin{subequations}\label{eq:LSM_edge}
\begin{eqnarray}
   \psi_{j\eta} & = & i C_\eta [-\text{sgn}(\mu)]^j \sinh (\eta j) \, , \\
   \phi_{j\eta} & = & i D_\eta [-\text{sgn}(\mu)]^{N+1-j} \sinh \big[ \eta(N \! + \! 1 \! - \! j) \big] \, , \quad
   \label{eq:LSM_phi_edge}
\end{eqnarray}
\end{subequations}
where $\text{sgn}(\cdot)$ is the sign function, and
the normalization constants $C_\eta, D_\eta$ are pure imaginary numbers satisfying
\begin{subequations}
  \begin{eqnarray}
    |C_\eta^2| & = & |D_\eta^2| \, , \\
    \dfrac{2}{|D_\eta|^2} & = & \dfrac{\sinh(\eta N)}{\sinh\eta} \cosh \big[ \eta(N+1) \big] - N  \, ,
    \quad \label{eq:normalization_edge}\\
    \dfrac{D_k}{C_k} & = & [-\text{sgn}(\mu)]^N \dfrac{|\mu| \sinh\eta}{\omega_\eta \sinh(\eta N)} \, .
    \label{eq:normalization_edge2}
\end{eqnarray}
\end{subequations}
Also in this case $C_\eta = \pm D_\eta$, while the relative sign depends on $\eta, \mu, N$
through Eqs.~\eqref{eq:normalization_edge},\eqref{eq:normalization_edge2}. The fact that
Eqs.~\eqref{eq:LSM_edge} are composed of real exponentials that decay from the borders
of the chain justifies our use of the term ``edge mode'', in relation to the complex-$k$ mode.

\subsection{Markovian master equation for a dissipative quadratic system}
\label{sec:MasterEq}

We now want to couple the Kitaev chain~\eqref{eq:kitaev} to a set of $N_B$ independent thermal baths,
each characterized by a temperature $T_n$ and a chemical potential $\mu_n$,
with $n \in \{1, \ldots, N_B\}$, and whose internal dynamics is described by a continuous model
of free fermions with Hamiltonian
\begin{equation}
   H_{\mathcal{E},n} = \int dq \: \epsilon_n(q) \, c_n^\dagger(q) c_n(q) \, .
\end{equation}
In general, we suppose that the $n$-th bath is coupled to a sublattice $\mathcal{I}_n$
of the chain through the linear interaction Hamiltonian
\begin{equation}
  \label{eq:interaction}
   H_{\rm int} = \sum_{n=1}^{N_B} \sum_{p \in \mathcal{I}_n} \int dq \, g_n(q)
   \Big( a_p + a_p^\dagger \Big) \Big[ c_n(q) + c_n^\dagger(q) \Big] \, .
\end{equation}

In this paper we will limit ourselves to the situation in which the baths relaxation times are much
smaller than the typical evolution time of the chain (more precisely, of its density operator
written in the interaction picture), thus making the use of a Born-Markov approximation well
justified~\cite{Petruccione-07}.
Furthermore we adopt a full secular approximation, whose regime of validity is discussed below
and explicitly calculated for typical situations in Appendix~\ref{appendixB}.
These hypotheses allow us to employ the self-consistent microscopic derivation
of Ref.~\cite{DAbbruzzo-21}, leading to the following Lindblad master equation
for the density operator $\rho_\mathcal{S}(t)$ (we assume units of $\hbar = k_B = 1$):
\begin{equation}
  \label{eq:ME}
  \frac{d\rho_\mathcal{S}(t)}{dt} = -i \big[ H_\mathcal{S} + H_{\rm LS}, \rho_\mathcal{S}(t) \big]
  + \mathcal{D}[\rho_\mathcal{S}(t)] \, ,
\end{equation}
provided the spectrum of the system $\{\omega_k\}$ is nondegenerate.
The dissipator $\mathcal{D}[\cdot]$ takes the form
\begin{multline}
   \mathcal{D}[\rho] = \sum_{n=1}^{N_B} \sum_k \gamma_{n,k} \Big[ (1-f_n(\omega_k))
   \left( 2b_k \rho b_k^\dagger - \{b_k^\dagger b_k, \rho\} \right) \\
   + f_n(\omega_k) \left( 2b_k^\dagger \rho b_k - \{b_k b_k^\dagger, \rho\} \right) \Big] \, ,
\end{multline}
with $f_n(\omega) = [1 + e^{(\omega - \mu_n)/T_n}]^{-1}$ being the Fermi-Dirac distribution associated
with the $n$-th bath. The coupling constants are
\begin{equation}
  \label{eq:couplings}
   \gamma_{n,k} = \mathcal{J}_n(\omega_k) \Big( \sum_{p\in\mathcal{I}_n} \phi_{pk} \Big)^2
   \equiv \mathcal{J}_n(\omega_k) \Phi_{n,k} \, ,
\end{equation}
where $\phi = A+B$ is the LSM matrix defined in the previous section,
while $\mathcal{J}_n(\omega) = \pi \int dq |g_n(q)|^2 \delta[\omega - \epsilon_n(q)]$
is the spectral density of the $n$-th bath. The Lamb-shift Hamiltonian takes the form
\begin{equation}
   H_{\rm LS} = \sum_k \lambda_k b_k^\dagger b_k \, ,
\end{equation}
where
\begin{equation}
   \lambda_k = \sum_{n=1}^{N_B} \frac{2\omega_k \Phi_{n,k}}{\pi} \:
          {\rm P} \!\! \int_0^\infty \! \frac{\mathcal{J}_n(\epsilon)}{\omega_k^2 - \epsilon^2} \, d\epsilon
          \label{eq:Lamb}
\end{equation}
and ${\rm P}$ stands for the sign of Cauchy's principal part.

The solution of the Lindblad master equation~\eqref{eq:ME} shows that the populations of the various
normal modes $\langle b_k^\dagger b_k \rangle$ evolve independently from one another
towards the steady state, each with a decay constant equal to $2 \sum_n \gamma_{n,k}$. When the latter
is nonzero for all $k$, the steady state is unique and is characterized by the
expectation values
\begin{equation}
  \label{eq:steadystate}
   \langle b_k^\dagger b_q \rangle_s = \delta_{kq} \, \dfrac{\sum_n \gamma_{n,k} f_n(\omega_k)}{\sum_n \gamma_{n,k}} \, .
\end{equation}
Concerning the nondiagonal terms of the density operator, in particular the expectation values
$\langle b_k^\dagger b_q \rangle$ and $\langle b_k^\dagger b_q^\dagger \rangle$ with $k \neq q$,
one can prove that they decay to zero with decay constant $\sum_n (\gamma_{n,k} + \gamma_{n,q})$
and with superimposed oscillations at frequencies of, respectively,
$\widetilde{\omega}_k \pm \widetilde{\omega}_q$, where $\widetilde{\omega}_k \equiv \omega_k + \lambda_k$.
In contrast, if for some pair $(k,q)$ one has $\sum_n \gamma_{n,k} = \sum_n \gamma_{n,q} = 0$,
the expectation values oscillate indefinitely around the initial values, with frequencies
$\widetilde{\omega}_k \pm \widetilde{\omega}_q$~\cite{DAbbruzzo-21}.

\subsection{Full secular approximation}
\label{sec:secular}

Before presenting our results for the dynamics of a dissipative Kitaev chain, as predicted
by the Lindblad master equation~\eqref{eq:ME}, it is worth stressing that such equation
requires the full secular approximation to hold, the validity of which has to be carefully assessed,
in order to avoid problems related to the debate between local and global master equations (see, e.g.,
Refs.~\cite{Rivas-10, Barra-15, Katz-16, Adesso-17, Hofer-17, Esposito-17, DeChiara-18, Cattaneo-19, DeChiara-20, Farina-20}).
In fact, to warrant it, one must require~\cite{Petruccione-07}
\begin{equation}
  \label{eq:secular_condition}
   \min_{k,q} |\omega_k - \omega_q| \gg \Gamma \, ,
\end{equation}
where $\Gamma$ is the inverse of the smallest typical evolution time of the density operator of the
chain $\widetilde{\rho}_\mathcal{S}(t)$, written in the interaction picture. In our case, the value
of $\Gamma$ is determined by the interplay between various quantities, such
as the coupling constants $\gamma_{n,k}$ and the Lamb-shift frequencies $|\lambda_k \pm \lambda_q|$.
A good estimate is provided by
\begin{equation}
   \Gamma \simeq \max \bigg[ 2 \max_k \sum_n \gamma_{n,k}, \; 2 \max_k |\lambda_k|, \;
     \max_{k,q} |\lambda_k - \lambda_q| \bigg] \, .
   \label{eq:Gamma}
\end{equation}

To make things concrete, let us consider a typical situation where the thermal baths are described
by Ohmic spectral densities with a high-frequency exponential cutoff,
\begin{equation}
  \label{eq:ohmic}
   \mathcal{J}_n(\omega) = \alpha_n \, \omega \, e^{-\omega/\omega_{c,n}} \, ,
\end{equation}
where $\alpha_n$ is the damping factor and $\omega_{c,n}$ is the cutoff frequency.
In Appendix~\ref{appendixB} we show that, in the context of this work, fixing the length of chain
$N$, it is always possible to choose the parameters $\alpha_n$ and $\omega_{c,n}$ in such a way to
guarantee the validity of the secular condition in Eq.~\eqref{eq:secular_condition}.
Therefore, in the following we will always implicitly assume to operate in this regime.


\section{Single bath: interference effects}
\label{sec:OneBath}

\begin{figure}
   \centering
   \includegraphics[width=\columnwidth]{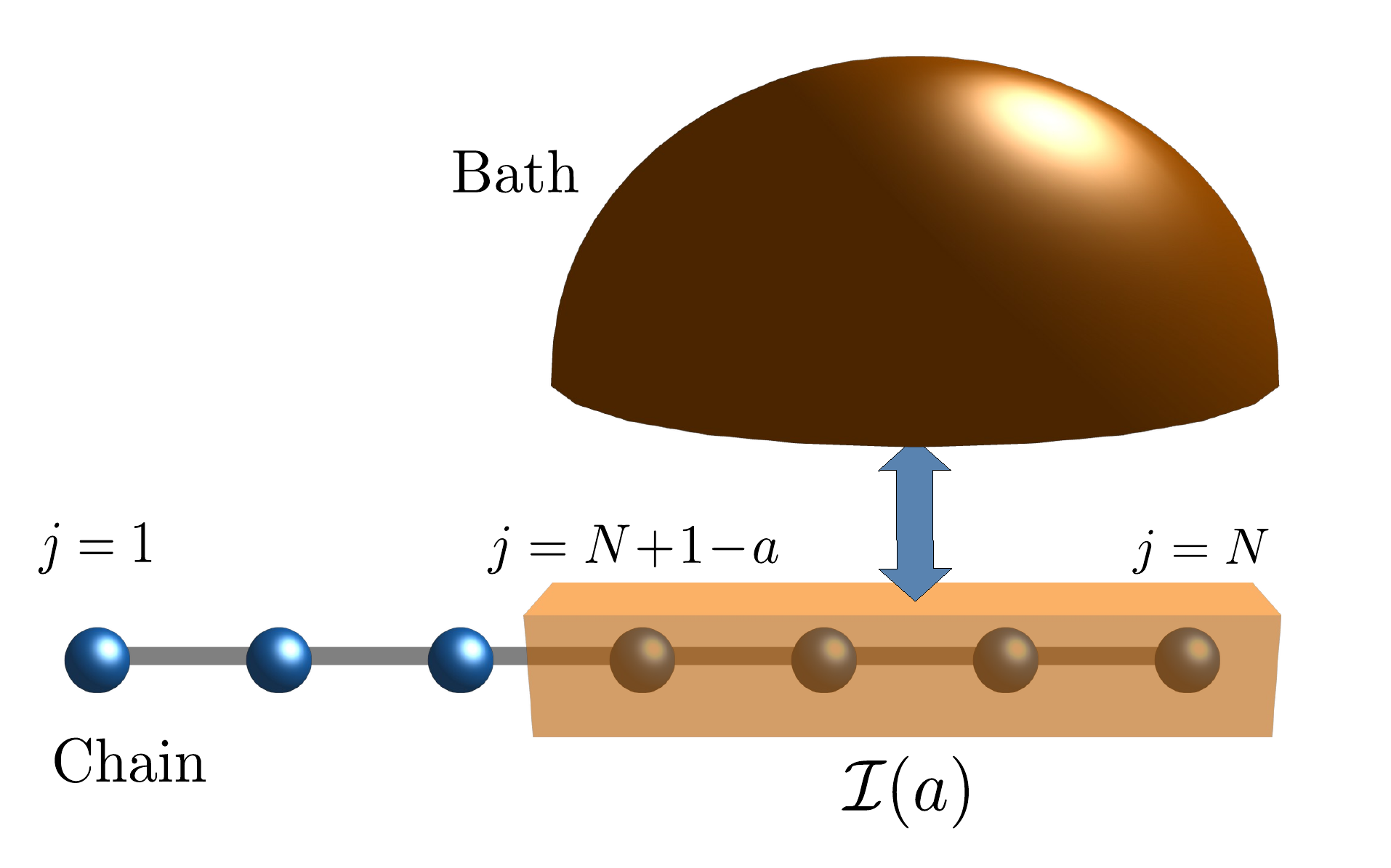}
   \caption{Single-bath setting. A Kitaev chain~\eqref{eq:kitaev} is coupled to a
   thermal bath through the linear interaction in Eq.~\eqref{eq:interaction}, where
   $\mathcal{I}(a)$ is the subchain involved in the coupling process.}
   \label{fig:system1}
\end{figure}

We first discuss the case of a single bath ($N_B = 1$),
thus dropping the subscript ${}_n$. The setting we have in mind is depicted in
Fig.~\ref{fig:system1}: a thermal bath is coupled to the chain through the interaction
Hamiltonian~\eqref{eq:interaction}, where $\mathcal{I} \equiv \mathcal{I}(a)$ is the subchain
identified by borders located in $j = N+1-a$ and $j=N$. The parameter $a \in \{1,\ldots,N\}$
represents the number of sites involved in the coupling process.

If the coupling is nontrivial, the chain eventually reaches a steady state which corresponds
to the thermal equilibrium with the bath, as predicted by Eq.~\eqref{eq:steadystate}.
Here it is interesting to analyze the relaxation time of each normal mode, which is related
to the coupling constants $\gamma_k$ defined in Eq.~\eqref{eq:couplings}.
In particular, once the environment spectral density is fixed [e.g., the Ohmic one,
written in Eq.~\eqref{eq:ohmic}], we need to calculate the quantity $\Phi_k$.
For bulk modes with real $k$, we can use the known expression of $\phi_{pk}$ from
Eq.~\eqref{eq:LSM_phi} to write
\begin{eqnarray}
  \label{eq:Phi1}
   \Phi_k &\equiv& \bigg[ \sum_{p\in\mathcal{I}(a)} \phi_{pk} \bigg]^2 =
   D_k^2 \bigg[ \sum_{p=N+1-a}^N \!\! \sin \big\{ k(N+1-p) \big\} \bigg]^2 \nonumber \\
   &=& D_k^2 \bigg[ \sum_{p=1}^a \sin(kp) \bigg]^2 \, .
\end{eqnarray}
Now, writing $\sin(kp) = {\rm Im}[e^{ikp}]$, we can interpret the summation in Eq.~\eqref{eq:Phi1}
as a geometric sum, which can be calculated explicitly. The result is
\begin{equation}
  \label{eq:Phi_bulk}
   \Phi_k = D_k^2 \, \frac{\sin^2(ka/2) \: \sin^2[k(a+1)/2]}{\sin^2(k/2)} \, .
\end{equation}
In contrast, for the complex-$k$ mode, we need to employ the formula in Eq.~\eqref{eq:LSM_phi_edge}
for the eigenvector expressed in terms of the imaginary part $\eta$.
Using a similar technique, it is possible to show that
\begin{subequations}
  \label{eq:Phi_edge}
\begin{eqnarray}
  \Phi^{(\mu < 0)}_\eta & = & |D_\eta|^2 \frac{\sinh^2(\eta a/2) \: \sinh^2(\eta(a+1)/2)}{\sinh^2(\eta/2)} \, ,\\
  \Phi^{(\mu > 0)}_\eta & = & |D_\eta|^2 \frac{\sinh^2(\eta a/2) \: \cosh^2(\eta(a+1)/2)}{\cosh^2(\eta/2)} \, , \quad
\end{eqnarray}
\end{subequations}
where we distinguished the two cases $\mu < 0$ and $\mu > 0$.

\begin{figure}
   \centering
   \includegraphics[width=\columnwidth]{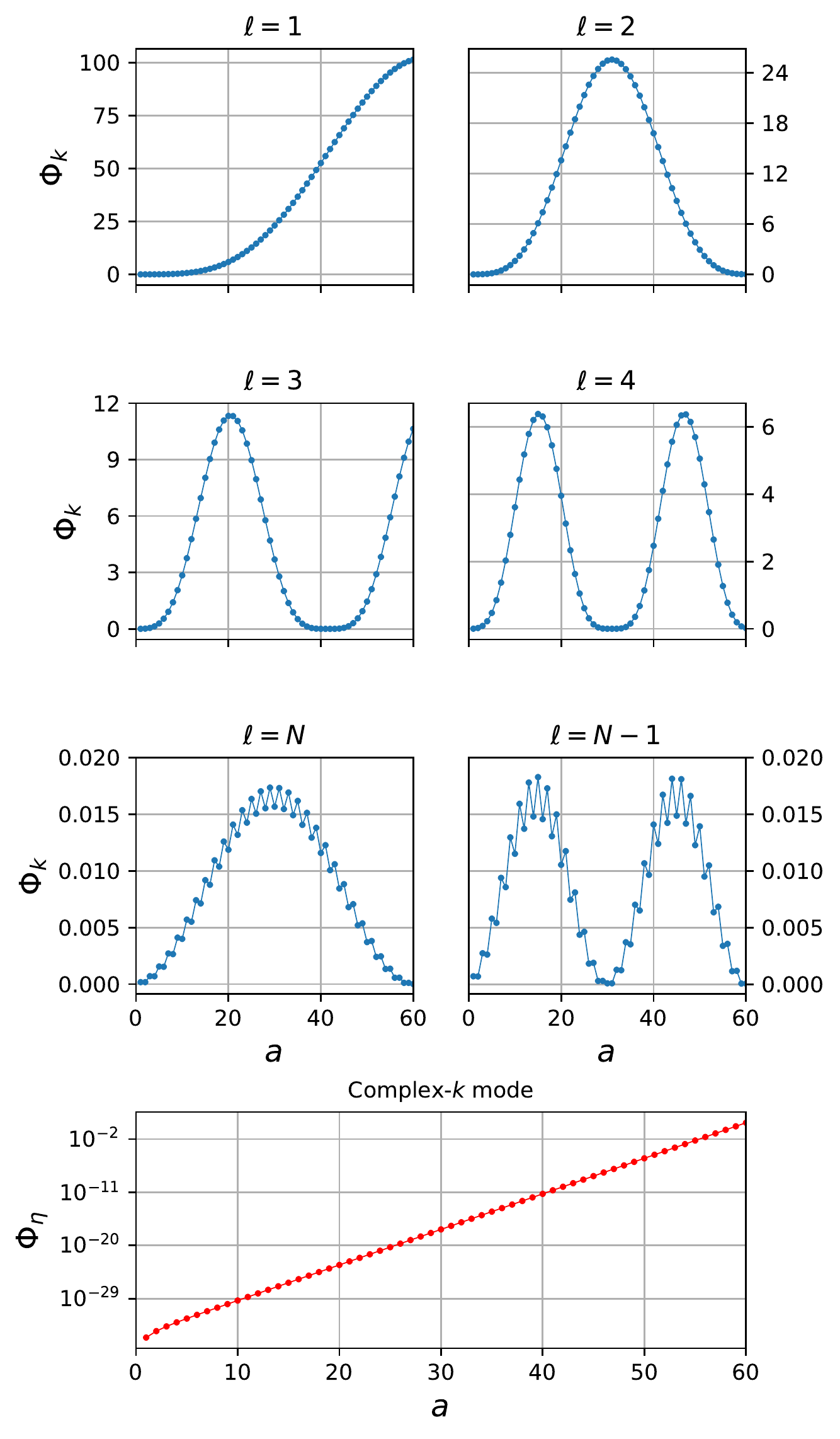}
   \caption{Blue dots (top panels): $\Phi_k$ in Eq.~\eqref{eq:Phi_bulk} as a function of the number $a$
     of sites coupled to the bath, for $\mu = -3$ and $k$ chosen in correspondence of the $\ell$-th
     bulk mode (the latter are ordered by increasing energy). Constructive and destructive
     interference patterns are due to the presence of multiple paths for the tunneling
     of fermions into and out of the system. Namely, for $\ell \leq \lfloor N/2 \rfloor$ there are
     $\lfloor \ell \rfloor$ interference peaks, while for larger $\ell$ beating structures appear.
     Red dots (bottom panel): $\Phi_\eta$ in Eqs.~\eqref{eq:Phi_edge} as a function of $a$,
     for $~{\mu = -1}$; the exponential increase with $a$ can be ascribed to the structure
     of the edge mode wavefunction in Eq.~\eqref{eq:LSM_phi_edge}. Here we fix $N=60$.
     Results are not significantly influenced by the specific value of $\mu$.}
   \label{fig:interference}
\end{figure}

At this point we would like to analyze the behavior of these quantities when changing the number of
sites $a$ coupled to the bath. To do that, we numerically calculate these expressions using the
technique described in Appendix~\ref{appendixA} [see Eq.~\eqref{eq:Phi_final}].
In Fig.~\ref{fig:interference} (upper panels) we report the results for $\Phi_k$ as a function of $a$
for some bulk modes of a Kitaev chain with $N=60$ sites and $\mu = -3$.
We observe that making $a$ greater does not necessarily lead to an increase in the chain-bath coupling:
in fact, we obtain curves that are typical of an interference pattern (except for
the lowest-energy mode, $\ell=1$). Analogous results are obtained with different values of $\mu$.
In the bottom panel of the same figure we also report the behavior of $\Phi_\eta$ for the complex-$k$ edge mode
when $\mu = -1$. This time we obtain an exponential increase of the coupling with $a$:
the effect can be easily traced back to the spatial structure of the wavefunction of the edge mode in
Eq.~\eqref{eq:LSM_phi_edge}, which is built with exponentials that decay starting from the borders of the chain.

The appearance of interference effects in the curves for $\Phi_k$ vs $a$, for the bulk modes of the chain,
can be intuitively explained by the fact that if the thermal bath is coupled to many sites,
a fermion can tunnel from a system to the other through multiple potential paths (see the schematics
in Fig.~\ref{fig:system1}). In quantum mechanics, it is well known that such a situation leads to
interference-related phenomena.
At a mathematical level, a more precise statement can be made by highlighting the link between the
coupling constants (in the form of $\Phi_k$) and the tunneling amplitudes. To this purpose, let us
define the quantity $\varphi_k$ by
\begin{equation}
   \Phi_k = \varphi_k^2 \, , \qquad
   \varphi_k = \sum_{p \in \mathcal{I}(a)} \phi_{pk} \, .
\end{equation}
It is easy to prove that $\varphi_k$ equals the probability amplitude associated with an
elementary process of population or depopulation of the normal mode with quantum number $k$, which
occurs by means of the interaction operator
$O = \sum_{p \in \mathcal{I}(a)} \left( a_p + a_p^\dagger \right)$ built from Eq.~\eqref{eq:interaction}.
To show that, we denote with $\ket{k} = b_k^\dagger \ket{\Omega}$ the eigenstate of the chain
obtained from an excitation of the quasiparticle vacuum $\ket{\Omega}$ with energy $\omega_k$.
The BV transformation~\eqref{eq:BV_direct} allows us to write
\begin{eqnarray}
   \mel{\Omega}{a_p}{k} & = & \! \sum_q \! \left( A_{pq} \mel{\Omega}{b_q b_k^\dagger}{\Omega} +
   B_{pq} \mel{\Omega}{b_q^\dagger b_k^\dagger}{\Omega} \right) = A_{pk}, \nonumber \\
   \mel{\Omega}{a_p^\dagger}{k} & = & \! \sum_q \! \left( A_{pq} \mel{\Omega}{b_q^\dagger b_k^\dagger}{\Omega} +
   B_{pq} \mel{\Omega}{b_q b_k^\dagger}{\Omega} \right) = B_{pk}. \nonumber
\end{eqnarray}
Using $\phi = A + B$, we obtain
\begin{equation}
   \mel{\Omega}{a_p + a_p^\dagger}{k} = \phi_{pk}
   \hspace{10pt} \Rightarrow \hspace{10pt}
   \mel{\Omega}{O}{k} = \varphi_k \, ,
\end{equation}
which is what we wanted to prove.


\section{Two baths: anomaly factor}
\label{sec:TwoBaths}

\begin{figure}
   \centering
   \includegraphics[width=\columnwidth]{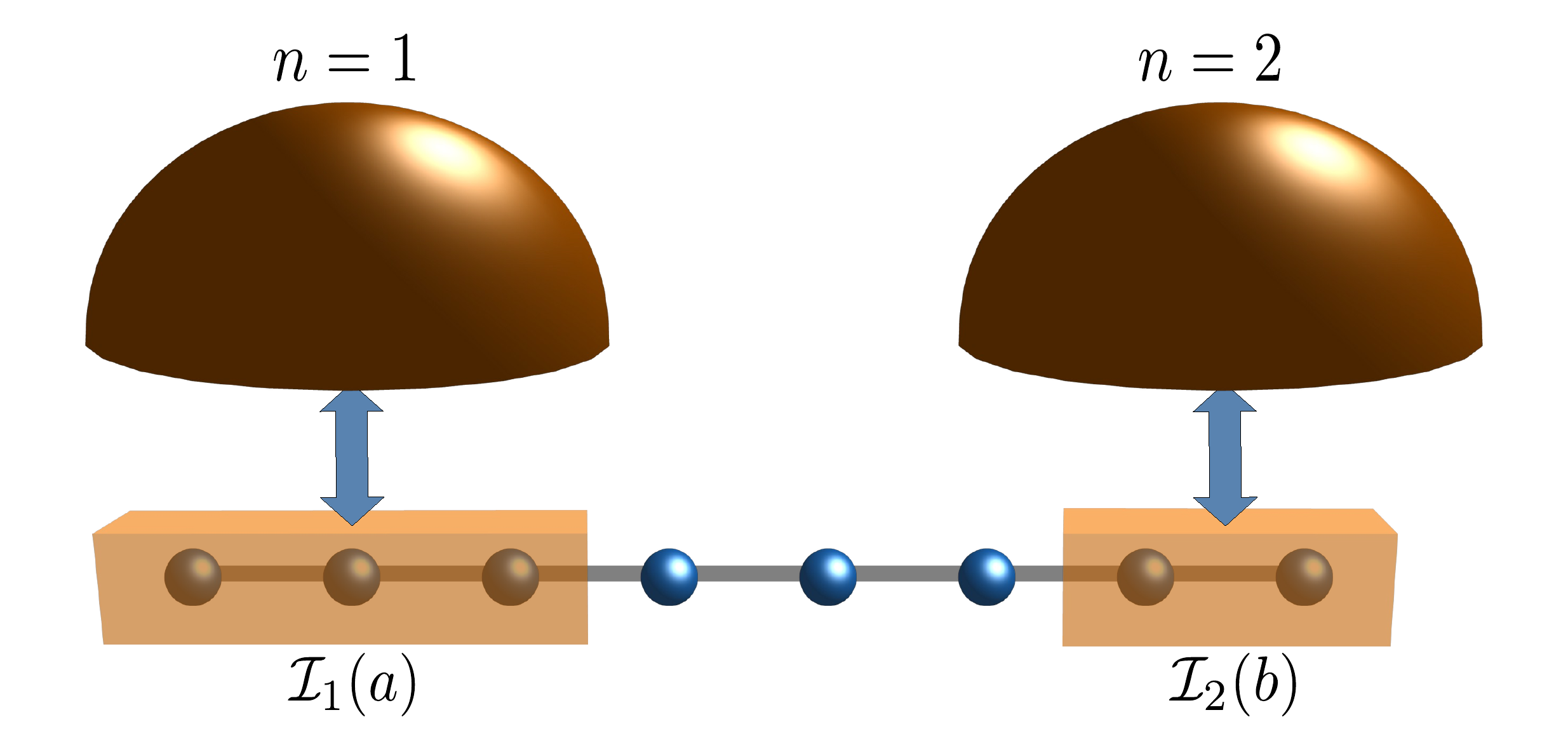}
   \caption{Two-bath setting for the study of the steady-state transport quantities.
     A Kitaev chain~\eqref{eq:kitaev} is coupled to two thermal baths through the linear interaction
     in Eq.~\eqref{eq:interaction}, where $\mathcal{I}_1(a)$ and $\mathcal{I}_2(b)$ are the subchains
     involved in the coupling process.}
   \label{fig:system2}
\end{figure}

Considering now $N_B = 2$, we address the setting depicted in Fig.~\ref{fig:system2}:
two thermal baths respectively interact with the subchains $\mathcal{I}_1(a)$ and $\mathcal{I}_2(b)$,
where $a$ and $b$ stand for the number of sites involved in the coupling, as before.
For what we are going to discuss below, it is not crucial to precisely fix the sites that compose
these subchains; however, for the sake of clarity, we suppose that the two baths
exert their influence on the edges of the chain, as in a standard transport measurement experiment.

If the two baths are characterized by different temperatures and/or chemical potentials,
steady-state currents may appear. In fact, starting from the master equation~\eqref{eq:ME}, one can obtain
a continuity equation for the average number of fermions in the chain, from which the steady-state particle
current can be calculated~\cite{DAbbruzzo-21}.
Using the notations introduced in Sec.~\ref{sec:MasterEq}, the current of particles
flowing from the reservoir $n=1$ into the system can be expressed as
\begin{equation}\label{eq:current}
  J = \sideset{}{'}\sum_{k} \frac{2S_k \gamma_{1,k}\gamma_{2,k}}{\gamma_{1,k} + \gamma_{2,k}}
  \Big[ f_1(\omega_k) - f_2(\omega_k) \Big] \, ,
\end{equation}
where the prime symbol constrains the sum over those $k$ such that $\gamma_{1,k},\gamma_{2,k} \neq 0$.
As noted in Ref.~\cite{DAbbruzzo-21}, this formula reduces to the well-known Landauer-B\"{u}ttiker
current, except for the fact that the transfer factor multiplying the difference
$f_1(\omega_k) - f_2(\omega_k)$ deviates from the standard expression for transport across multiple
ballistic channels~\cite{Datta} by a factor $S_k$ that is defined in terms of the LSM matrices as
\begin{equation}
  \label{eq:anomaly}
   S_k = \big( A^T A - B^T B \big)_{kk} = \sum_{j=1}^N \phi_{jk} \psi_{jk} \, .
\end{equation}
The emergence of this quantity is directly related to the presence of anomalous
non-number-conserving terms in the system Hamiltonian. Indeed, if those terms were absent,
it would have been possible to arrange the BV transformation~\eqref{eq:BV_direct}
in such a way to have $B = 0$ and $A$ unitary. In this situation one would have obtained
$S_k \equiv 1$, recovering the standard transfer factor. For this reason,
we will refer to $S_k$ as the {\it anomaly factor}.

It is not immediate to infer general properties from the definition in Eq.~\eqref{eq:anomaly}.
The only obvious structural feature comes from the constraint $A^T A + B^T B = 1$,
which implies that $(A^T A)_{kk}$ and $(B^T B)_{kk}$ can only take values in the closed interval $[0,1]$
(given also the fact that $A^T A$ and $B^T B$ are positive semidefinite matrices).
This means that $S_k \in [-1,1]$.
Notice how, in principle, $S_k$ can assume negative values,
hence bringing a negative contribution to the particle conductance.

\begin{figure}
   \centering
   \includegraphics[width=\columnwidth]{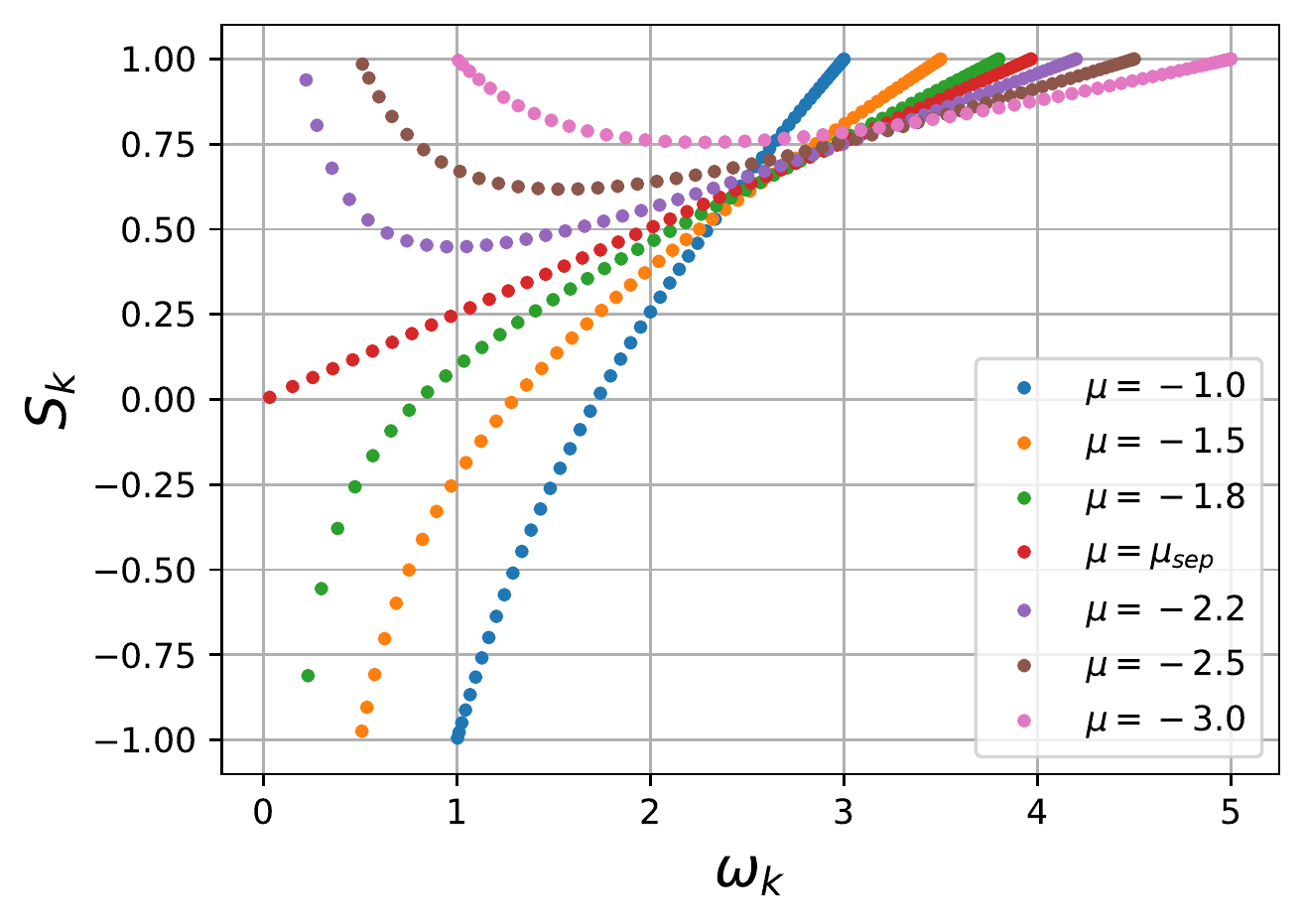}
   \caption{The anomaly factor in Eq.~\eqref{eq:anomaly_bulk} for the bulk modes
     as a function of $\omega_k$, for different values of $\mu<0$ and $N=60$.
     The separation point between the regions with and without the complex-$k$ edge mode is located
     at $\mu_{\rm sep} \equiv -2N/(N+1)$. The curves have a very different shape in the two
     cases $\mu > \mu_{\rm sep}$ and $\mu < \mu_{\rm sep}$, signaling that $S_k$
     contains information about the finite-size analogue of the Pfaffian topological
     invariant of the chain [see Eq.~\eqref{eq:anomaly_bound}].}
   \label{fig:anomaly}
\end{figure}

In the present context, we can explicitly calculate the anomaly factor for
the Kitaev chain, using the known expressions for the LSM matrices.
For what concerns the bulk modes, using Eqs.~\eqref{eq:LSM} we write
\begin{eqnarray}
   S_k &=& C_k D_k \sum_{j=1}^N \sin (kj) \: \sin \! \big[ k(N+1-j) \big] \nonumber \\
   &=& \frac{C_k D_k}{2} \left\{ \frac{\sin kN}{\sin k} - N \cos \big[ k(N+1) \big] \right\} \, .
\end{eqnarray}
Now, writing $C_k$ in terms of $D_k$ by means of Eq.~\eqref{eq:normalization2}, expanding
$\cos[k(N+1)]$ with the addition formula and, finally, using Eq.~\eqref{eq:sine_rewrite} to
rewrite the ratio $\sin kN/\sin k$, we have that
\begin{equation}
   S_k = -\frac{D_k^2}{2\omega_k} \Big[ 2N\cos k + \mu(N+1) \Big] \, .
\end{equation}
This expression can be rewritten for convenience as follows, using $|\mu_{\rm sep}| = 2N/(N+1)$:
\begin{subequations}
\label{eq:anomaly_formulas}
\begin{equation}
   \label{eq:anomaly_bulk}
   S_k = -\frac{D_k^2 N}{\omega_k} \left( \cos k + \frac{\mu}{|\mu_{\rm sep}|} \right) \, .
\end{equation}
A similar procedure can be followed starting from Eqs.~\eqref{eq:LSM_edge} to calculate the anomaly
factor associated with the complex-$k$ mode, obtaining
\begin{equation}
  S_\eta = -\frac{|D_\eta|^2 N}{\omega_\eta} \left( \cosh \eta - \frac{|\mu|}{|\mu_{\rm sep}|} \right) \, .
     \label{eq:anomaly_edge}
\end{equation}
\end{subequations}
These quantities can be numerically calculated using the same technique as the one employed in
Sec.~\ref{sec:OneBath} and described in Appendix~\ref{appendixA}.
It turns out that the anomaly factor of the complex-$k$ mode in Eq.~\eqref{eq:anomaly_edge} is
practically zero for sufficiently large $N$, therefore in the following we focus on the anomaly
factor of the bulk modes in Eq.~\eqref{eq:anomaly_bulk}, which shows a nontrivial behavior.
Figure~\ref{fig:anomaly} reports some curves of $S_k$ vs $\omega_k$ for different values of $\mu$,
here taken as a negative number. We can observe how the shape of $S_k$
drastically changes according to whether the complex-$k$ mode is present or not in the spectrum.
In particular, if the complex-$k$ mode is absent, we observe the existence of a positive lower
bound, so that $S_k$ is shifted towards the unit value. On the other hand,
if the complex-$k$ mode is present, $S_k$ can assume values all over the interval $[-1,1]$.
In particular, the negative conductance phenomenon indeed appears in this situation.

We argue that this behavior is due to the fact that $S_k$ contains information about the finite-size
analogue of a topological invariant of the chain. Intuitively speaking, from Eq.~\eqref{eq:anomaly_bulk}
we have that, for $\mu < 0$,
\begin{equation}
  \label{eq:anomaly_bound}
   S_k \geq \frac{D_k^2 N}{\omega_k} \left( \frac{|\mu|}{|\mu_{\rm sep}|} - 1 \right)
   = \frac{D_k^2 N}{\omega_k} \left| \frac{|\mu|}{|\mu_{\rm sep}|} - 1 \right| \mathcal{W} \, ,
\end{equation}
where
\begin{equation}
  \mathcal{W} = \text{sgn}( |\mu/\mu_{\rm sep}| - 1 ) \, .
  \label{eq:Wtopo}
\end{equation}
If $|\mu| > |\mu_{\rm sep}|$ (i.e., the edge mode is absent) then $~{\mathcal{W} = +1}$
and a positive lower bound on the anomaly factor emerges.
The larger $|\mu|$ the bigger the bound is, until the curve is completely pushed
towards the maximum allowed value $S_k = 1$. On the other hand, if $|\mu| < |\mu_{\rm sep}|$
(i.e., the edge mode is present) then $\mathcal{W} = -1$ and the lower bound becomes negative.
If $|\mu|$ is close enough to zero, the presence of the bound amounts to say that $S_k$ is larger
than the minimum allowed value, $S_k \geq -1$. It is worth noticing that $\mathcal{W}$
can be viewed as the finite-size analogue of the Pfaffian topological invariant~\cite{KaneMele},
which is obtained by substituting $|\mu_{\rm sep}| \rightarrow |\mu_c| = 2$ in Eq.~\eqref{eq:Wtopo}.

A more precise analytical characterization can be made in the thermodynamic limit.
Here the correspondence with the microscopic model described by the master
equation~\eqref{eq:ME} is generally lost, since the secular approximation cannot be regarded as
justified (see Appendix~\ref{appendixB}). However this idealized situation is still useful to
understand the behavior of the finite-size case.

Indeed, as $N \rightarrow \infty$, we can approximate $|\mu_{\rm sep}| \simeq 2$ and also assume
that $k$ is a continuous variable in the closed interval $[0,\pi]$.
In this case, it is convenient to look at the $N$-independent quantity
\begin{equation}
  \label{eq:renorm_anomaly}
  R(k) = \frac{S_k}{D_k^2 N} = - \frac{\cos k + \mu/2}{\sqrt{\mu^2 + 4\mu \cos k + 4}} \, .
\end{equation}
In fact, when $N$ is large, the quantity $D_k^2 N$ becomes a constant independent from $k$:
\begin{equation}
   \frac{1}{D_k^2 N} = \frac{N+1}{2N} - \frac{\cos^2 kN}{N(2+\mu\cos k)}
   \xrightarrow{N \rightarrow \infty} \frac{1}{2}
\end{equation}
[Eq.~\eqref{eq:norm_cheby} was used to rewrite the normalization constant].
As a consequence, in the thermodynamic limit, the $R(k)$ is just a rescaled anomaly factor,
keeping all the properties of $S_k$. A straightforward calculation of the derivative $dR/dk$
shows that $R(k)$ always admits two stationary points at the boundaries $k=0,\pi$.
An additional stationary point can appear in the interior of the domain, for $k=k_s$, where one finds
\begin{equation}
   R(k_s) = -\frac{1}{2\mu} \frac{\mu^2 - 4}{\sqrt{\mu^2 - 4}} \, .
\end{equation}
Notice that this point exists only in the trivial phase, where $\mu^2 > 4$.
Moreover, the second derivative $d^2 R/dk^2$ shows that, when present, it is a positive minimum
(negative maximum) for $\mu < 0$ ($\mu > 0$). In particular, we recover the presence
of a positive lower bound observed in Fig.~\ref{fig:anomaly} for negative chemical potentials.
On the other hand, in the topological phase $\mu^2 < 4$, the maximum and the minimum are necessarily
reached at the boundaries, since $R(k)$ is a continuous function of $k$ over a compact domain.
Inspecting $R(0)$ and $R(\pi)$, we find that the image of $S_k$ is necessarily the
entire interval $[-1,1]$, in agreement with what has been observed in the finite-size situation.

\begin{figure}
   \centering
   \includegraphics[width=\columnwidth]{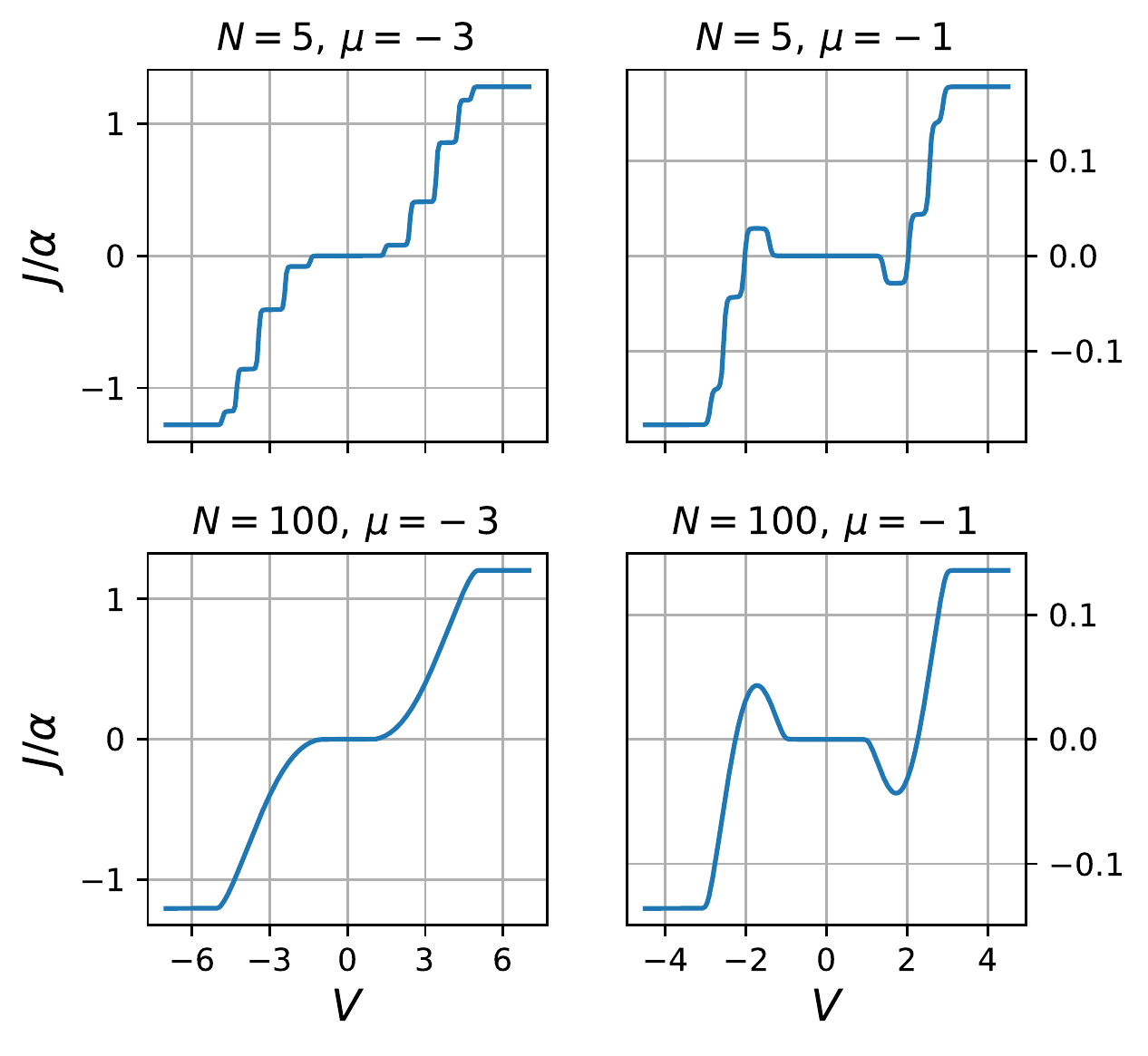}
   \caption{Particle current (in units of damping factor $\alpha$) as a function of the
   potential difference between the two baths, for different values of $N$ and $\mu$.
   The value of $V$ on the $x$-axis is chosen such that $\mu_1 = V$ and $\mu_2 = -V$, while the temperature for
   both baths is fixed at $T = 0.03$. The cutoff frequency is fixed at $\omega_c = 5$.
   If $N$ is small enough (upper panels), we clearly observe step-like changes of the current,
   occurring when $|V|$ matches an eigenenergy of the chain, while for larger values of $N$
   the curves become smoother.
   Also note that, for $|\mu| > |\mu_{\rm sep}|$ (left panels, absence of edge modes)
   the current grows monotonically with $V$, while for $|\mu| < |\mu_{\rm sep}|$ (right panels,
   presence of an edge mode) the behavior of ${J=J(V)}$ becomes nonmonotonic.}
   \label{fig:current}
\end{figure}

We finally point out that the behavior of the anomaly factor qualitatively affects
that of the particle current in Eq.~\eqref{eq:current}, which may be directly observed in experiments.
Assuming to have baths differing only by their chemical potentials, Fig.~\ref{fig:current} reports
some examples of currents as functions of the potential difference.
Notice that, for sufficiently small chain lengths (upper panels), a step-like behavior
of the current $J$ vs. the bias $V$ emerges, witnessing a resonance induced by a given
eigenenergy of the chain. On the other hand, for larger chains (lower panels) such
steps are less visible.
Most interestingly, the shape of the curve turns out to be sensibly different according to
whether the complex-$k$ mode is present or not (i.e., depending on the phase of the system,
in the thermodynamic limit), as expected from the analysis of the anomaly factor reported
in Fig.~\ref{fig:anomaly}: in the presence of an edge mode, the particle current may
change nonmonotonically with the bias, thus signaling the possibility to have a negative conductance.


\section{Conclusions and outlook}
\label{sec:Conclusions}

We unveiled how the topological features of the Kitaev chain reflect in its quantum transport properties,
when the system is locally coupled to two thermal baths held at different temperatures.
Specifically, we showed that the fermionic particle current flowing into the system resembles
that of the Landauer-B{\"u}ttiker scattering formalism, with the additional presence of an anomaly
factor keeping track of the zero-temperature phase of the system.
Our results are fully consistent with a microscopically derived model, which keeps into account
the Markovian nature of the system-bath coupling, provided a full secular approximation is enforced.

In the future, it would be interesting to extend such calculations to non-quadratic
  (interacting) Hamiltonian models or to models which
keep into account the possibility to have arbitrarily large system-environment interactions,
where the open-system dynamics becomes inevitably non-Markovian and may be affected by important
memory effects~\cite{Rivas-14, Breuer-16, deVega}.
The latter represents the typical scenario in realistic solid-state situations,
where a more direct comparison with experimentally feasible situations would be possible.

\acknowledgments
The authors would like to thank G.~Piccitto and M.~Polini for fruitful discussions.


\appendix

\section{Diagonalization of a finite Kitaev chain with open boundary conditions}
\label{appendixA}

In this appendix we provide details on how to apply the LSM method~\cite{LSM-61}
to the diagonalization problem of the open Kitaev chain in Eq.~\eqref{eq:general_kitaev},
with $t = \Delta$ and $\mu \neq 0$. Although this is a quite standard problem, in particular in the
context of spin chains, it is not immediate to find a properly developed treatment of this system
at finite size, in the literature (see, however,
Refs.~\cite{Kao-14, Zvyagin-15, Hegde-15, Zeng-19, Leumer-20, Leumer-21}).
Here we only discuss the case with $t = \Delta$ and $\mu \neq 0$, cf. Eq.~\eqref{eq:kitaev},
where calculations can be made simpler.

\subsection{Lieb-Schultz-Mattis method}

Let us start with a brief reminder on the LSM method for the diagonalization of a fermionic quadratic
Hamiltonian with real coefficients, generally written as
\begin{equation}
   \label{eq:quadratic_hamiltonian}
   H_\mathcal{S} = \sum_{i,j=1}^N \left[ Q_{ij} a_i^\dagger a_j + \tfrac{1}{2} P_{ij} \big(
   a_i^\dagger a_j^\dagger + a_j a_i \big) \right] \, .
\end{equation}
The starting point consists in defining new fermionic operators via a BV transformation,
of the form
\begin{equation}
   \label{eq:BV_inverse}
   b_k = \sum_{j=1}^N \left( X_{kj} a_j + Y_{kj} a_j^\dagger \right) \, .
\end{equation}
Note that this is the inverse transformation of the one in Eq.~\eqref{eq:BV_direct},
thus $X = A^T$ and $Y = B^T$.

The LSM insight resides in the fact that if the transformation~\eqref{eq:BV_inverse} allows to
write the Hamiltonian~\eqref{eq:quadratic_hamiltonian} as
$H_\mathcal{S} = \sum_k \omega_k b_k^\dagger b_k + \text{const.}$, then
it must be true that $[b_k, H_\mathcal{S}] = \omega_k b_k$. By writing this relation in terms of the $a_j$
one can obtain the following consistency equations in terms of the vectors
${\bf X}_k^T = (X_{k1}, \ldots, X_{kN})$ and ${\bf Y}_k^T = (Y_{k1}, \ldots, Y_{kN})$:
\begin{subequations}
\begin{eqnarray}
   Q {\bf X}_k + P {\bf Y}_k &=& \omega_k {\bf X}_k \, , \\
   -Q {\bf X}_k - P {\bf Y}_k &=& \omega_k {\bf Y}_k \, .
\end{eqnarray}
\end{subequations}
In terms of new vectors ${\bm \phi}_k = {\bf X}_k + {\bf Y}_k$ and
${\bm \psi}_k = {\bf X}_k - {\bf Y}_k$ we can rewrite
\begin{subequations}
\label{eq:coupled_LSM}
\begin{eqnarray}
   (Q+P) {\bm \phi}_k &=& \omega_k {\bm \psi}_k \, , \label{eq:coupled_LSM1} \\
   (Q-P) {\bm \psi}_k &=& \omega_k {\bm \phi}_k \, . \label{eq:coupled_LSM2}
\end{eqnarray}
\end{subequations}
We will refer to these as the coupled LSM equations. A decoupled form can be obtained by multiplying
to the left the first one by $(Q-P)$, and the second one by $(Q+P)$:
\begin{subequations}
  \label{eq:eigen_problem}
\begin{eqnarray}
   V {\bm \phi}_k &=& \omega_k^2 {\bm \phi}_k \, , \\
   W {\bm \psi}_k &=& \omega_k^2 {\bm \psi}_k \, , \label{eq:eigen_problem_W}
\end{eqnarray}
\end{subequations}
where $V \equiv (Q-P)(Q+P)$ and $W \equiv (Q+P)(Q-P)$.
This corresponds to a standard eigenvalue problem; by solving it, one finds both
the excitation spectrum $\{\omega_k\}$ and the BV coefficients. Notice also that, for example,
\begin{equation}
   ({\bm \phi}_k)_j = X_{kj} + Y_{kj} = A_{jk} + B_{jk} = \phi_{jk} \, ,
\end{equation}
hence the normalized eigenvectors found with this procedure are the same quantities entering
our Lindblad master equation~\eqref{eq:ME}.

\subsection{Application to the Kitaev chain}

We now focus on the specialized Kitaev Hamiltonian written in Eq.~\eqref{eq:kitaev}.
The matrices $Q$ and $P$ defined in Eq.~\eqref{eq:quadratic_hamiltonian} become
\begin{equation}
   Q = \left(
   \begin{array}{cccc}
      -\mu & -1 & & \\
      -1 & -\mu & -1 & \\
      & -1 & -\mu & \ddots \\
      & & \ddots & \ddots
   \end{array}
   \right) , \hspace{5pt}
   P = \left(
   \begin{array}{cccc}
      0 & -1 & & \\
      1 & 0 & -1 & \\
      & 1 & 0 & \ddots \\
      & & \ddots & \ddots
   \end{array}
   \right) ,
\end{equation}
from which we can build the matrices involved in the eigenvalue problem of
Eqs.~\eqref{eq:eigen_problem},
\begin{subequations}
\begin{eqnarray}
   V &=& \left(
   \begin{array}{ccccc}
      \mu^2 & 2\mu & & & \\
      2\mu & 4+\mu^2 & 2\mu & & \\
      & 2\mu & 4+\mu^2 & \ddots & \\
      & & \ddots & \ddots & 2\mu \\
      & & & 2\mu & 4+\mu^2
   \end{array}
   \right) , \\
   W &=& \left(
   \begin{array}{ccccc}
      4+\mu^2 & 2\mu & & & \\
      2\mu & \ddots & \ddots & & \\
      & \ddots & 4+\mu^2 & 2\mu & \\
      & & 2\mu & 4+\mu^2 & 2\mu \\
      & & & 2\mu & \mu^2
   \end{array}
   \right) . \label{eq:W}
\end{eqnarray}
\end{subequations}
From the shape of these matrices, we can infer that ${\bm \phi}_k$ must be the same as ${\bm \psi}_k$,
modulo an inversion of the indexes $j \rightarrow N+1-j$. Therefore, it is enough to solve for
${\bm \psi}_k$, for example, by finding eigenvalues and eigenvectors of $W$. The eigenvalue
equation~\eqref{eq:eigen_problem_W} translates in the following system:
\begin{align}
  \label{eq:eigen_system}
    2\mu ({\bm \psi}_k)_{j-1} + (4+\mu^2) ({\bm \psi}_k)_j + 2\mu ({\bm \psi}_k)_{j+1}& =
   \omega_k^2 ({\bm \psi}_k)_j \, ,\nonumber\\
    (4+\mu^2) ({\bm \psi}_k)_1 + 2\mu ({\bm \psi}_k)_2 &= \omega_k^2 ({\bm \psi}_k)_1 \, ,\nonumber\\
    2\mu ({\bm \psi}_k)_{N-1} + \mu^2 ({\bm \psi}_k)_N &= \omega_k^2 ({\bm \psi}_k)_N \, , \quad
\end{align}
where the first equation is defined in the bulk, $~{2 \leq j \leq N-1}$, and the others assume the
role of boundary conditions. Since we have a translational invariant bulk, it is reasonable to guess
an ansatz solution of the form
\begin{equation}
   \label{eq:ansatz}
   ({\bm \psi}_k)_j = -\frac{iC_k}{2} \left( e^{ikj} + \alpha_k e^{-ikj} \right) \, ,
\end{equation}
where $k$ assumes the role of a quantum number (in general, a complex one), while $C_k$, $\alpha_k$
are parameters to be determined. The global multiplicative constant is chosen in a format that will
turn out to be convenient later.

By substituting~\eqref{eq:ansatz} in the first condition of the system~\eqref{eq:eigen_system}, we can find the functional
form of the eigenvalues given in Eq.~\eqref{eq:spectrum},
\begin{equation}
  \label{eq:spectrum_app}
   \omega_k^2 = \mu^2 + 4\mu \cos k + 4 \, .
\end{equation}
To ensure that $\omega_k^2$ is a real nonnegative number, we have to impose some constraints on $k$.
Namely, splitting real and imaginary part as $k = \kappa + i \eta$, we can rewrite
\begin{equation}
   \omega_k^2 = \mu^2 + 2\mu \big[ (e^\eta + e^{-\eta})\cos\kappa + i(e^\eta - e^{-\eta})\sin\kappa \big] + 4 \, ,
\end{equation}
and thus there are only two possibilities:
\begin{itemize}
\item $\eta = 0$, that is, $k$ is real. In this case, given the periodicity of
  Eq.~\eqref{eq:spectrum_app}, we can limit ourselves to $k \in [0,\pi]$;

\item $\sin\kappa = 0$, that is, either $\kappa = 0$ or $\kappa = \pi$.
  In this case, the eigenvalue can be expressed in terms of the imaginary part $\eta$ only
  as $\omega_\eta^2 = \mu^2 \pm 4\mu \cosh\eta + 4$ (the $+$ sign is for $\kappa = 0$,
  the $-$ sign is for $\kappa = \pi$).
\end{itemize}

If we now substitute the ansatz~\eqref{eq:ansatz} in the second condition of the
system~\eqref{eq:eigen_system} and use the expression in Eq.~\eqref{eq:spectrum_app} for
$\omega_k^2$, it is easy to find out that $\alpha_k = -1$. This means that
\begin{equation}
  \label{eq:LSM_psi_app}
   ({\bm \psi}_k)_j = C_k \sin (kj) \, ,
\end{equation}
Notice that, in the case of real $k$, we must now rule out
the possibilities $k = 0$ and $k = \pi$, since they would lead to ${\bm \psi}_k = 0$.

Finally, the third condition in the system~\eqref{eq:eigen_system} provides us with
the quantization condition on the quantum number $k$, which turns out to be
\begin{equation}
  \label{eq:quantization_app}
   2 \sin (kN) + \mu \sin [ k(N+1) ] = 0 \, ,
\end{equation}
as anticipated in Eq.~\eqref{eq:quantization}. An analysis of this condition is crucial to understand
the finite-size effects induced by the QPT of the model. Let us fix $N$ and define the function
\begin{equation}
  \label{eq:F}
   F(k) \equiv -\frac{2 \sin kN}{\sin [ k(N+1) ]} \, .
\end{equation}
The condition~\eqref{eq:quantization_app} means that it must be $F(k) = \mu$,
for every allowed $k$.

\begin{figure}
   \centering
   \includegraphics[width=\columnwidth]{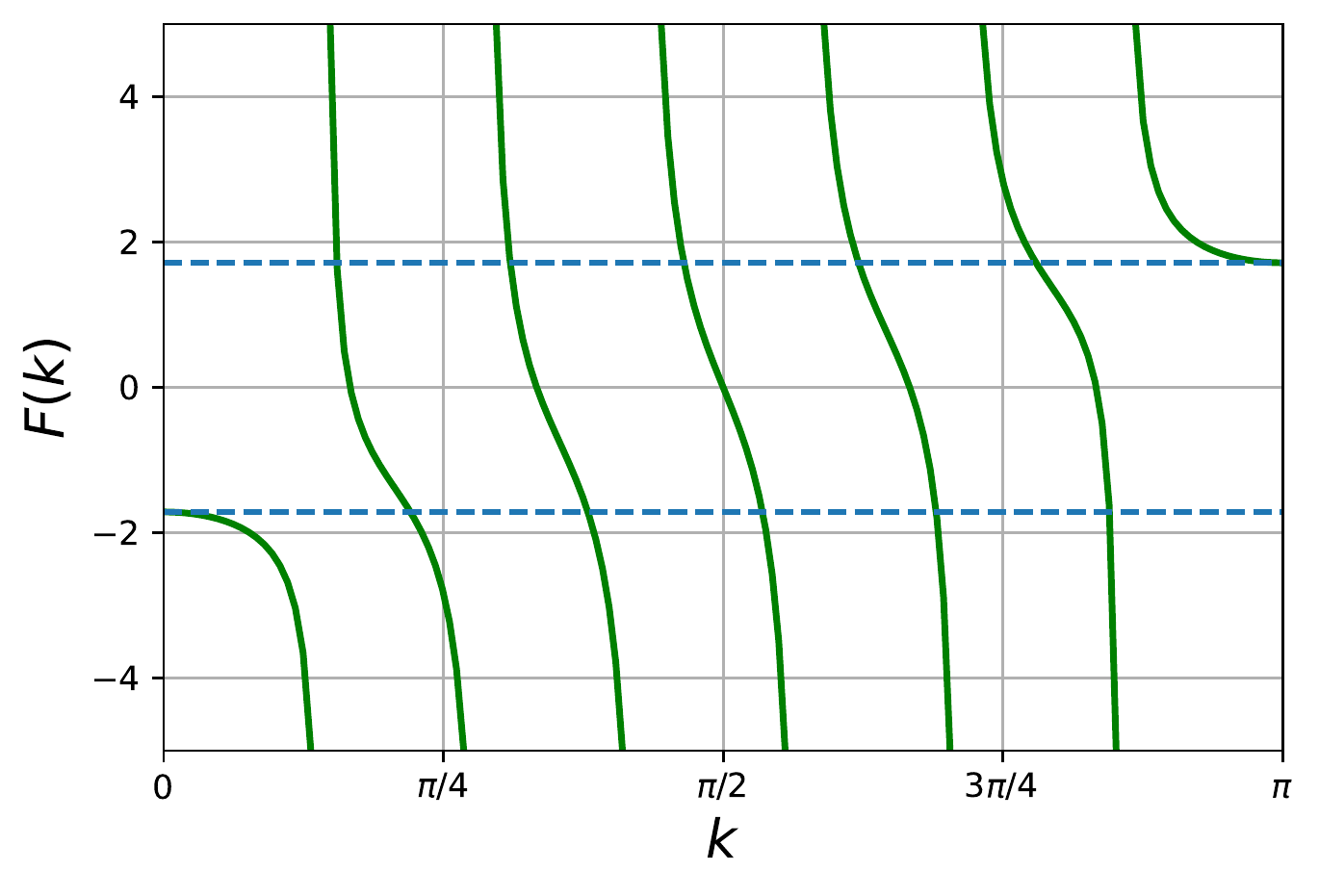}
   \caption{The function $F(k)$ in Eq.~\eqref{eq:F}, in the case of real $k$ and
   for fixed $N=6$. We can see the existence of two separate regions in which the equation
   $F(k) = \mu$ admits, respectively, $N$ solutions (when $|\mu| > |\mu_{\rm sep}|$) and $N-1$
   solutions (when $|\mu| < |\mu_{\rm sep}|$). In the latter case, the missing solution has to be
   found in the complex domain and constitutes the finite-size analogue of the zero-energy
   Majorana edge mode. The blue dashed lines indicate the separation lines $\pm |\mu_{\rm sep}|$.}
   \label{fig:quantization}
\end{figure}

To study this equation, in Fig.~\ref{fig:quantization} we plot
the function $F(k)$ at fixed $N$, in the case where $k \in (0,\pi)$ is a real number.
We can see that there are $N$ real solutions if $|\mu|$ is bigger than a certain threshold value
$|\mu_{\rm sep}|$, while under this value there are only $N-1$ real solutions. It is also simple to
verify that this behavior does not depend on the particular choice of $N$. The crossing between the two
situations happens when $\mu$ is equal to one of the stationary points of $F(k)$ located in $k=0$
and $k=\pi$. Thus, we infer that
\begin{equation}
  \label{eq:separation_app}
   |\mu_{\rm sep}| = \lim_{k \rightarrow \pi} F(k) = \frac{2N}{N+1} \, .
\end{equation}
When $|\mu| < |\mu_{\rm sep}|$ the missing solution has to be found in the complex domain.
In particular, if $\mu < 0$ one has to impose $\kappa = 0$, while if $\mu > 0$ one has to impose
$\kappa = \pi$, in accordance with the general discussion made before on the reality of
$\omega_k^2$. The value of the imaginary part $\eta$ is then fixed by the quantization condition.
The formulas~\eqref{eq:spectrum_app}, \eqref{eq:LSM_psi_app}, and~\eqref{eq:quantization_app}
can be respectively rewritten for this complex-$k$ mode as
\begin{subequations}
\begin{eqnarray}
   \omega_\eta^2 & = & \mu^2 - 4|\mu| \cosh \eta + 4 \, ,\\
   ({\bm \psi}_\eta)_j & = & iC_\eta \big[ -\text{sgn}(\mu)\big]^j \sinh (\eta j) \, , \label{eq:complex_LSM_app}\\
   0 & = & 2\sinh (\eta N) - |\mu| \sinh \big[\eta(N+1)\big] \, . \quad \label{eq:complex_quant_app}
\end{eqnarray}
\end{subequations}
Notice that, if $\eta$ satisfies Eq.~\eqref{eq:complex_quant_app}, then $-\eta$ satisfies it as well,
therefore we can safely assume $\eta > 0$ (the energy $\omega_\eta$ does not change and the eigenvectors
only acquire an overall irrelevant minus sign).

We claim that this mode constitutes the finite-size analogue of the zero-energy Majorana edge mode,
which is properly obtained only in the thermodynamic limit. To show this, let us first split
the hyperbolic sine in Eq.~\eqref{eq:complex_quant_app} by using the formula for the sine of a sum.
This allows us to rewrite the quantization condition as
\begin{equation}
   \label{eq:rewrite_quanti_complex_app}
   \frac{\sinh \eta}{\sinh (\eta N)} = \frac{2 - |\mu| \cosh \eta}{|\mu| \cosh(\eta N)} \, .
\end{equation}
Using this expression and $~{\cosh^2 \eta = 1 + \sinh^2 \eta}$, we get
\begin{equation}
   \frac{\cosh^2 \eta}{\cosh^2(\eta N)} + \frac{4}{|\mu|} \tanh^2 (\eta N) \left( \cosh\eta
   - \frac{1}{|\mu|} \right) - 1 = 0 \, .
\end{equation}
In the thermodynamic limit $N \rightarrow \infty$, we remain with
\begin{equation}
   \cosh\eta_\infty = \frac{\mu^2 + 4}{4|\mu|} \hspace{10pt} \Rightarrow \hspace{10pt}
   \omega^2_{\eta_\infty} = 0 \, ,
\end{equation}
which is what we wanted to prove.

The parameter $C_k$ in Eq.~\eqref{eq:LSM_psi_app} can be found by imposing a normalization condition
$\sum_j ({\bm \psi}_k)_j^2 = 1$. It is easy to show that, if $k$ is real,
$C_k$ must be taken real as well, and satisfying
\begin{equation}
   \frac{2}{C_k^2} = N - \frac{\sin kN}{\sin k} \cos \big[ k(N+1) \big] \,.
\end{equation}
On the other hand, for the complex-$k$ mode, we must have a pure imaginary $C_\eta$
(in order to consistently have a real LSM eigenvector) with
\begin{equation}
   \frac{2}{|C_\eta|^2} = \frac{\sinh(\eta N)}{\sinh\eta} \cosh \big[ \eta(N+1) \big] - N \, .
\end{equation}

According to our previous observation on the symmetry by indexes inversion, we can write the
eigenvectors of $V$ right away as
\begin{equation}
   ({\bm \phi}_k)_j = D_k \, \sin \big[ k(N+1-j) \big] \, ,
\end{equation}
with an analogue expression to Eq.~\eqref{eq:complex_LSM_app} for the complex-$k$ mode.
Here $D_k$ is a normalization constant which generally differs from $C_k$. Actually, since
$\sum_j \sin^2(kj) = \sum_j \sin^2[k(N+1-j)]$ we can infer that $|D_k|^2 = |C_k|^2$, therefore the
ratio $D_k/C_k$ must have modulus one. Namely, there exists a function $s(k)$ such that
$|s(k)|=1$ and
\begin{equation}
   D_k = s(k) \, C_k \, .
\end{equation}
To determine $s(k)$, we  go back to the coupled LSM equations~\eqref{eq:coupled_LSM}.
In particular, writing the first component of Eq.~\eqref{eq:coupled_LSM2}, we obtain
\begin{equation}
   s(k) = -\frac{\mu \sin k}{\omega_k \sin (kN)} \, ,
\end{equation}
For the complex-$k$ mode, this expression can be written in terms of the imaginary part $\eta$ as
\begin{equation}
   s(\eta) = \big[ -\text{sgn}(\mu) \big]^N \frac{|\mu| \sinh\eta}{\omega_\eta \sinh(\eta N)} \, .
\end{equation}
In both cases $s$ is real, hence $s(k) = \pm 1$ and the normalization
constants $C_k,D_k$ differ, at most, by a sign.


\subsection{Numerical calculations}

In Secs.~\ref{sec:OneBath} and \ref{sec:TwoBaths} we used the results obtained from the
diagonalization procedure to calculate relevant quantities of the model,
namely the couplings constants [see Eqs.~\eqref{eq:Phi_bulk} and~\eqref{eq:Phi_edge}]
and the anomaly factor [see Eqs.~\eqref{eq:anomaly_formulas}].
In practice, to plot these quantities, a naive approach would be to find the quantum
numbers $k$ (or $\eta$) by numerically solving the quantization condition~\eqref{eq:quantization},
which is a nontrivial transcendental equation.

Here we show that it is possible to follow an alternative and more convenient strategy, based on the
following observation. If one is able to write the quantities of interest in such a way to make $k$
appear only in the form of $(\cos k)$, then Eq.~\eqref{eq:spectrum} would allow us to rewrite the
formulas in terms of the spectrum $\{\omega_k\}$, thus removing the explicit dependence on $k$.
The advantage relies on the fact that, from a numerical point of view, the spectrum is easily
obtained through a simple diagonalization procedure: it is in fact enough to diagonalize the
tridiagonal real symmetric matrix $W$ defined in Eq.~\eqref{eq:W}, rather than solving a
transcendental equation.

For example, let us rewrite Eqs.~\eqref{eq:Phi_bulk} and~\eqref{eq:Phi_edge} according to this
insight. We employ the Chebyshev polynomial of the first kind $T_n(x)$, defined as the unique
polynomial satisfying~\cite{Abramowitz}
\begin{equation}
   T_n(x) = \Bigg\{
   \begin{array}{ll}
      \cos \big[ n \arccos x \big] & |x| \leq 1 \,, \\
      \cosh \big[ n \, \text{arccosh} \, x \big] & x \geq 1\,, \\
      (-1)^n \cosh \big[ n \, \text{arccosh}(-x) \big] \; & x \leq -1 \, .
   \end{array}
\end{equation}
In our case, we need the relations $T_n(\cos k) = \cos(nk)$ and $T_n(\cosh \eta) = \cosh(n\eta)$
(remember that $\eta$ is assumed to be positive). Equation~\eqref{eq:Phi_bulk} becomes
\begin{subequations}\label{eq:Phi_final}
\begin{equation}
   \Phi_k = D_k^2 \frac{[1-T_a(\cos k)] \: [1-T_{a+1}(\cos k)]}{1-\cos k} \, ,
\end{equation}
while Eqs.~\eqref{eq:Phi_edge} can be grouped into
\begin{equation}
   \Phi_\eta = |D_\eta|^2 \frac{[T_a(\cosh\eta)-1] \: [T_{a+1}(\cosh\eta) + \text{sgn}(\mu)]}{ \cosh\eta + \text{sgn}(\mu) } \, .
\end{equation}
\end{subequations}
Notice that the Chebyshev polynomial can be efficiently calculated by any standard numerical library.
To express the normalization constants $D_k, D_\eta$ according to this scheme,
it is sufficient to use the sine addition formula and rewrite the quantization
condition~\eqref{eq:quantization} as
\begin{equation}
   \frac{\sin (kN)}{\sin k} = -\frac{\mu \cos (kN)}{2 + \mu \cos k}
   \label{eq:sine_rewrite}
\end{equation}
[see Eq.~\eqref{eq:rewrite_quanti_complex_app} for its counterpart with complex $k$].
Substituting this into Eqs.~\eqref{eq:normalization} and~\eqref{eq:normalization_edge},
and expanding the cosines, we get
\begin{subequations}
\label{eq:norm_cheby}
\begin{eqnarray}
   \frac{1}{D_k^2} & = & \frac{N+1}{2} - \frac{T_N^2(\cos k)}{2+\mu\cos k} \, , \\
   \frac{1}{|D_\eta|^2} & = & \frac{T_N^2(\cosh \eta)}{2-|\mu|\cosh\eta} - \frac{N+1}{2} \, ,
\end{eqnarray}
\end{subequations}
which conclude the sequence of formulas we used to generate the plots in Fig.~\ref{fig:interference}.

The same reasoning can be applied for the anomaly factor $S_k$. Eqs.~\eqref{eq:anomaly_formulas}
are already in the desired shape, provided the normalization constants $D_k^2$ and $|D_\eta|^2$
are expressed according to Eqs.~\eqref{eq:norm_cheby}.


\section{Validity of the secular approximation}
\label{appendixB}

\begin{figure}
   \centering
   \includegraphics[width=\columnwidth]{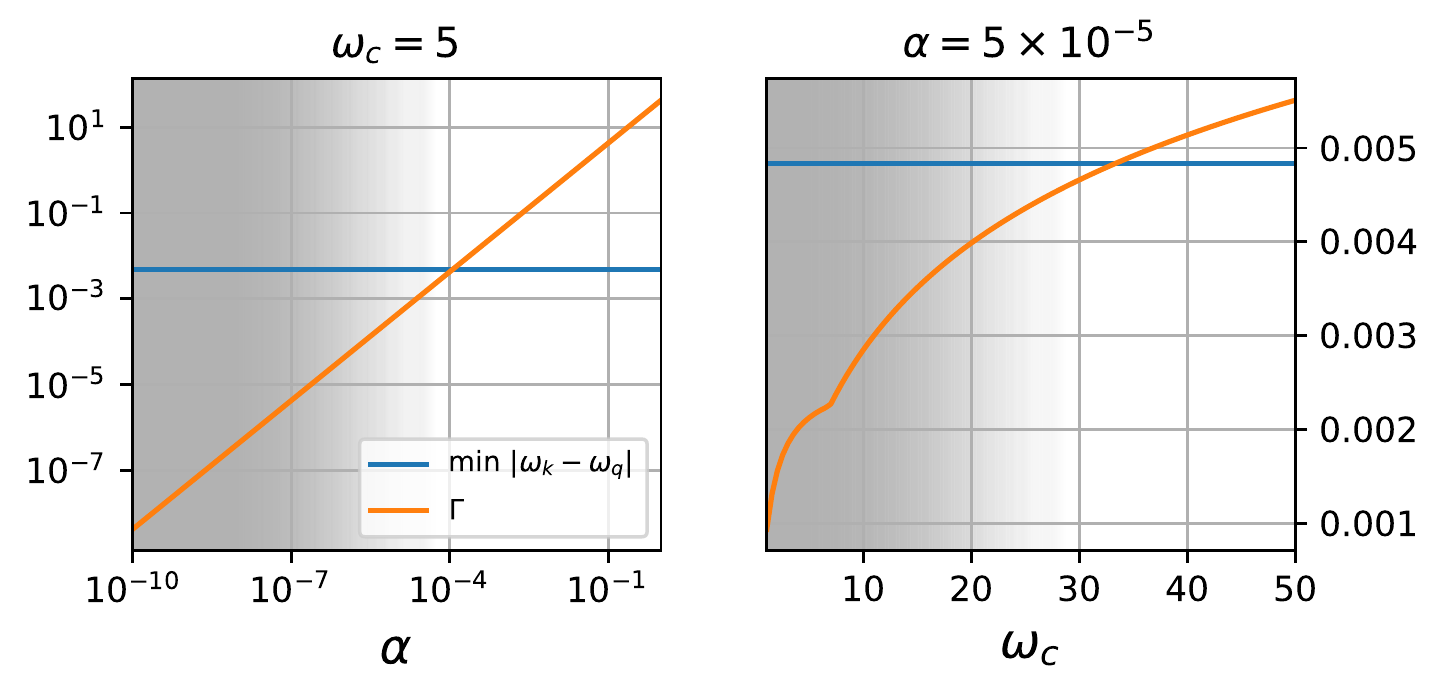}
   \caption{Comparison between the quantities involved in the secular
     condition~\eqref{eq:secular_condition}, as functions of $\alpha$ and $\omega_c$ for fixed
     $~{N=60}$, $\mu = -3$, and $a = N/2$. The blue horizontal line indicates the value of
     $\min |\omega_k - \omega_q|$, while the orange line is the plot of $\Gamma$.
     The shaded area denotes the parameter region where $\min |\omega_k-\omega_q| \gtrsim \Gamma$,
     and thus the secular approximation can be safely assumed.}
   \label{fig:fixedN}
   \vspace{10pt}
   \includegraphics[width=\columnwidth]{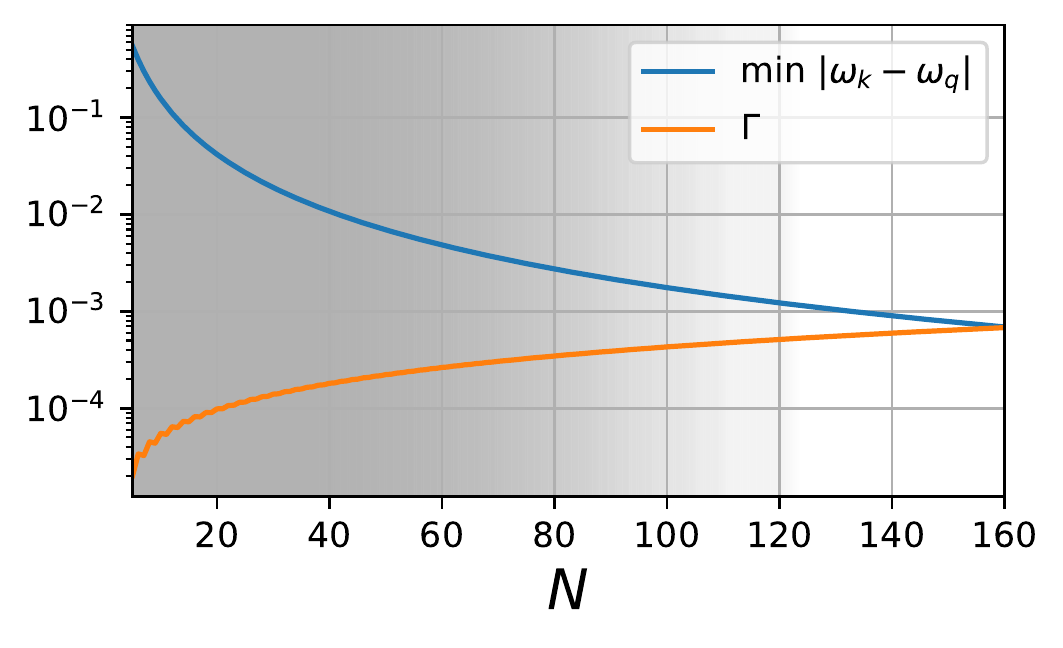}
   \caption{Same kind of analysis as in Fig.~\ref{fig:fixedN}, but fixing $\mu = -3$, $a = N/2$,
     $\alpha = 10^{-5}$, $\omega_c = 1.5$, and varying the number $N$ of sites.
     While it is always possible to lower the value of $\Gamma$ by altering the bath parameters,
     at fixed bath model the secular approximation turns out to be less justified at large $N$.}
   \label{fig:variableN}
\end{figure}

In Sec.~\ref{sec:MasterEq} we introduced a microscopically-derived Markovian master equation for the
dissipative Kitaev chain using the results of Ref.~\cite{DAbbruzzo-21}, and we used it for all the
subsequent reasoning. We also specified that Eq.~\eqref{eq:ME} can be considered valid only under
the hypothesis of full secular approximation, summarized by the inequality~\eqref{eq:secular_condition}.
In this appendix we justify the validity of that condition
in the context of the model under investigation. For simplicity here we only focus on
the single-bath case (see Fig.~\ref{fig:system1}), even if
our arguments below can be easily extended to multi-bath situations.

To this purpose, we need to evaluate the quantities entering the definition of the constant $\Gamma$,
cf.~Eq.~\eqref{eq:Gamma}, once assuming an Ohmic spectral density~\eqref{eq:ohmic} for the bath.
The coupling constants $\gamma_k = \mathcal{J}(\omega_k)\Phi_k$ have been calculated
in Sec.~\ref{sec:OneBath} [see, e.g., Eqs.~\eqref{eq:Phi_final}].
Concerning the Lamb-shift frequencies $\lambda_k$, we
notice that the integral in Eq.~\eqref{eq:Lamb} can be explicitly written in terms of the
exponential integral function ${\rm Ei}(x) = \int_{-\infty}^x dt \, e^t/t$, as
\begin{equation}
   \lambda_k = \frac{\alpha \omega_k \Phi_k}{\pi} \left[ e^{-\omega_k/\omega_c}
   {\rm Ei} \left( \frac{\omega_k}{\omega_c} \right)
   + e^{\omega_k/\omega_c} {\rm Ei} \left( -\frac{\omega_k}{\omega_c} \right) \right] .
\end{equation}
With these formulas at hand, we can numerically calculate $\Gamma$ and verify the validity
of the secular condition~\eqref{eq:secular_condition}.

Let us first look at the behavior of the involved quantities in a scenario with fixed $N$, $\mu$,
and $a$ (where $a$ is the number of sites coupled to the bath). In Fig.~\ref{fig:fixedN} we report
an example of results as functions of the bath parameters $\alpha$ and $\omega_c$,
fixing $\omega_c$ and $\alpha$, respectively. We observe that it is always possible
to choose a bath model (namely, to suitably select the bath parameters $\alpha$ and $\omega_c$)
in such a way that the secular approximation is justified to arbitrary degree.
A qualitatively analogous scenario emerges for any other values of the parameters $N$, $\mu$, $a$.

On the other hand, we can first choose a particular bath model, fixing $\alpha$ and $\omega_c$,
and then analyze what happens by varying the chain length $N$. An example of such analysis is reported in
Fig.~\ref{fig:variableN} (a qualitatively analogous behavior emerges for other values
of the bath parameters): from this we understand that, at fixed bath model, the secular approximation
turns out to be less justified for high enough values of $N$. This is a consequence of the fact that
the spectrum of the Kitaev chain becomes denser by increasing the number of sites (i.e., the minimum
distance between different eigenenergies tends to zero), while $\Gamma$ still acquires
nontrivial values. In particular, the secular condition is always violated in the thermodynamic
limit $N \rightarrow \infty$, at least assuming a nontrivial bath model with $\alpha \neq 0$.



\begin{thebibliography}{100}

\bibitem{Petruccione-07}
  H.-P. Breuer and F. Petruccione,
  \textit{The Theory of Open Quantum Systems}
  (Oxford University Press, Oxford, 2007).

\bibitem{Alicki-07}
  R. Alicki and K. Lendi,
  \textit{Quantum Dynamical Semigroups and Applications}
  (Springer-Verlag, Berlin Heidelberg, 2007).

\bibitem{Lindblad-76}
  G. Lindblad,
  On the generators of quantum dynamical semigroups,
  Commun. Math. Phys. {\bf 48}, 119 (1976);
  V. Gorini, A. Kossakowski, and E. C. G. Sudarshan,
  Completely positive dynamical semigroups of $N$-level systems,
  J. Math. Phys. {\bf 17}, 821 (1976).

\bibitem{Harbola-06}
  U. Harbola, M. Esposito, and S. Mukamel,
  Quantum master equation for electron transport through quantum dots and single molecules,
  Phys. Rev. B {\bf 74}, 235309 (2006).

\bibitem{Santos-16}
  J. P. Santos and G. T. Landi,
  Microscopic theory of a nonequilibrium open bosonic chain,
  Phys. Rev. E {\bf 94}, 062143 (2016).

\bibitem{Benatti-20}
  F. Benatti, R. Floreanini, and L. Memarzadeh,
  Bath-assisted transport in a three-site spin chain: Global versus local approach,
  Phys. Rev. A {\bf 102}, 042219 (2020).

\bibitem{Cattaneo-20}
  M. Cattaneo, G. L. Giorgi, S. Maniscalco, and R. Zambrini,
  Symmetry and block structure of the Liouvillian superoperator in partial secular approximation,
  Phys. Rev. A {\bf 101}, 042108 (2020).

\bibitem{Dorn-21}
  G. Dorn, E. Arrigoni, and W. von der Linden,
  Efficient energy resolved quantum master equation for transport calculations in large strongly correlated systems,
  J. Phys. A: Math. Theor. {\bf 54}, 075301 (2021).

\bibitem{DAbbruzzo-21}
  A. D'Abbruzzo and D. Rossini,
  Self-consistent microscopic derivation of Markovian master equations for open quadratic quantum systems,
  Phys. Rev. A {\bf 103}, 052209 (2021).

\bibitem{Sieberer-16}
  L. M. Sieberer, M. Buchhold, and S. Diehl,
  Keldysh field theory for driven open quantum systems,
  Rep. Prog. Phys. {\bf 79}, 096001 (2016).

\bibitem{Muller-12}
  M. M{\"u}ller, S. Diehl, G. Pupillo, and P. Zoller,
  Engineered open systems and quantum simulations with atoms and ions,
  Adv. At. Mol. Opt. Phys. {\bf 61}, 1 (2012).

\bibitem{Kitaev-01}
  A. Y. Kitaev,
  Unpaired Majorana fermions in quantum wires,
  Phys. Usp. {\bf 44}, 131 (2001).

\bibitem{DasSarma-10}
  R. M. Lutchyn, J. D. Sau, and S. Das Sarma,
  Majorana fermions and a topological phase transition in semiconductor-superconductor heterostructures,
  Phys. Rev. Lett. {\bf 105}, 077001 (2010).

\bibitem{vonOppen-10}
  Y. Oreg, G. Refael, and F. von Oppen,
  Helical liquids and Majorana bound states in quantum wires,
  Phys. Rev. Lett. {\bf 105}, 177002 (2010).

\bibitem{Alicea}
  J. Alicea,
  New directions in the pursuit of Majorana fermions in solid state systems,
  Rep. Prog. Phys. {\bf 75}, 076501 (2012).

\bibitem{Franz}
  M. Franz,
  Majorana's wires,
  Nat. Nanotechnol. {\bf 8}, 149 (2013).

\bibitem{Mourik-12}
  V. Mourik, K. Zuo, S. M. Frolov, S. R. Plissard, E. P. A. M. Bakkers, and L. P. Kouwenhoven,
  Signatures of Majorana fermions in hybrid superconductor-semiconductor nanowire devices,
  Science {\bf 336}, 1003 (2012).

\bibitem{Zhang-19}
  H. Zhang, D. E. Liu, M. Wimmer, and L. P. Kouwenhoven,
  Next steps of quantum transport in Majorana nanowire devices,
  Nat. Commun. {\bf 10}, 5128 (2019).

\bibitem{Nayak-08}
  C. Nayak, S. H. Simon, A. Stern, M. Freedman, and S. Das Sarma,
  Non-Abelian anyons and topological quantum computation,
  Rev. Mod. Phys. {\bf 80}, 1083 (2008).

\bibitem{RV-21}
  D. Rossini and E. Vicari,
  Coherent and dissipative dynamics at quantum phase transitions,
  Phys. Rep. (2021, in press),
  DOI: 10.1016/j.physrep.2021.08.003.

\bibitem{DasSarma-17}
  C.-X. Liu, J. D. Sau, and S. Das Sarma,
  Role of dissipation in realistic Majorana nanowires,
  Phys. Rev. B {\bf 95}, 054502 (2017).

\bibitem{Ng-15}
  H. T. Ng,
  Decoherence of interacting Majorana modes,
  Sci. Rep. {\bf 5}, 12530 (2015).

\bibitem{Molignini-17}
  P. Molignini, E. van Nieuwenburg, and R. Chitra,
  Sensing Floquet-Majorana fermions via heat transfer,
  Phys. Rev. B {\bf 96}, 125144 (2017).

\bibitem{Huang-19}
  Y. Huang, A. M. Lobos, and Zi Cai,
  Dissipative Majorana quantum wires,
  iScience {\bf 21}, 241 (2019).

\bibitem{vanCaspel-19}
  M. T. van Caspel, S. E. Tapias Arze, and I. P{\'e}rez Castillo,
  Dynamical signatures of topological order in the driven-dissipative Kitaev chain,
  SciPost. Phys. {\bf 6}, 026 (2019).

\bibitem{Dutta-20}
  S. Bandyopadhyay, S. Bhattacharjee, and A. Dutta,
  Dynamical generation of Majorana edge correlations in a ramped Kitaev chain coupled to nonthermal dissipative channels,
  Phys. Rev. B {\bf 101}, 104307 (2020).

\bibitem{Cooper-20}
  S. Lieu, M. McGinley, and N. R. Cooper,
  Tenfold way for quadratic Lindbladians,
  Phys. Rev. Lett. {\bf 124}, 040401 (2020).

\bibitem{Cooper-21}
  S. Lieu, M. McGinley, O. Shtanko, N. R. Cooper, and A. V. Gorshkov,
  Kramers' degeneracy for open systems in thermal equilibrium,
  arXiv:2105.02888 (2021).

\bibitem{Delgado-12}
  O. Viyuela, A. Rivas, and M. A. Martin-Delgado,
  Thermal instability of protected end states in a one-dimensional topological insulator,
  Phys. Rev. B {\bf 86}, 155140 (2012).

\bibitem{Delgado-14}
  O. Viyuela, A. Rivas, and M. A. Martin-Delgado,
  Uhlmann phase as a topological measure for one-dimensional fermion systems,
  Phys. Rev. Lett. {\bf 112}, 130401 (2014).

\bibitem{Kao-14}
  H.-c. Kao,
  Chiral zero modes in superconducting nanowires with Dresselhaus spin-orbit coupling,
  Phys. Rev. B {\bf 90}, 245435 (2014).

\bibitem{Zvyagin-15}
  A. A. Zvyagin,
  Majorana bound states in the finite-length chain,
  Low Temp. Phys. {\bf 41}, 625 (2015).

\bibitem{Hegde-15}
  S. Hegde, V. Shivamoggi, S. Vishveshwara, and D. Sen,
  Quench dynamics and parity blocking in Majorana wires,
  New J. Phys. {\bf 17}, 053036 (2015).

\bibitem{Zeng-19}
  C. Zeng, C. Moore, A. M. Rao, T. D. Stanescu, and S. Tewari,
  Analytical solution of the finite-length Kitaev chain coupled to a quantum dot,
  Phys. Rev. B {\bf 99}, 094523 (2019).

\bibitem{Leumer-20}
  N. Leumer, M. Marganska, B. Muralidharan, and M. Grifoni,
  Exact eigenvectors and eigenvalues of the finite Kitaev chain and its topological properties,
  J. Phys.: Condens. Matter {\bf 32}, 445502 (2020).

\bibitem{Leumer-21}
  N. Leumer, M. Grifoni, B. Muralidharan, and M. Marganska,
  Linear and nonlinear transport across a finite Kitaev chain: An exact analytical study,
  Phys. Rev. B {\bf 103}, 165432 (2021).

\bibitem{Pfeuty-70}
  P. Pfeuty,
  The one-dimensional Ising model with a transverse field,
  Ann. Phys. {\bf 57} 79 (1970).

\bibitem{Sachdev}
  S. Sachdev,
  \textit{Quantum phase transitions, 2nd ed.}
  (Cambridge University Press, Cambridge, 2011).

\bibitem{Blaizot-Ripka}
  J.-P. Blaizot and G. Ripka,
  \textit{Quantum Theory of Finite Systems} (MIT Press, 1986).

\bibitem{Xiao-09}
  M. Xiao,
  Theory of transformation for the diagonalization of quadratic Hamiltonians, arXiv:0908.0787.

\bibitem{LSM-61}
  E. Lieb, T. Schultz, and D. Mattis,
  Two soluble models of an antiferromagnetic chain,
  Ann. Phys. {\bf 16}, 407 (1961).

\bibitem{Rivas-10}
  {\'A}. Rivas, A. D. K. Plato, S. F. Huelga, and M. B. Plenio,
  Markovian master equations: A critical study,
  New J. Phys. {\bf 12}, 113032 (2010).

\bibitem{Barra-15}
  F. Barra, The thermodynamic cost of driving quantum systems by their boundaries,
  Sci. Rep. {\bf 5}, 14873 (2015).

\bibitem{Katz-16}
  G. Katz and R. Kosloff,
  Quantum thermodynamics in strong coupling: Heat transport and refrigeration,
  Entropy {\bf 18}, 186 (2016).

\bibitem{Adesso-17}
  J. O. González, L. A. Correa, G. Nocerino, J. P. Palao, D. Alonso, and G. Adesso,
  Testing the validity of the ‘local’ and ‘global’ GKLS master equations on an exactly solvable model,
  Open Syst. Inf. Dyn. {\bf 24}, 1740010 (2017).

\bibitem{Hofer-17}
  P. P. Hofer, M. Perarnau-Llobet, L. D. M. Miranda, G. Haack, R. Silva, J. B. Brask, and N. Brunner,
  Markovian master equations for quantum thermal machines: local versus global approach,
  New J. Phys. {\bf 19}, 123037 (2017),

\bibitem{Esposito-17}
  P. Strasberg, G. Schaller, T. Brandes, and M. Esposito,
  Quantum and information thermodynamics: A unifying framework based on repeated interactions,
  Phys. Rev. X {\bf 7}, 021003 (2017).

\bibitem{DeChiara-18}
  G. De Chiara, G. Landi, A. Hewgill, B. Reid, A. Ferraro, A. J. Roncaglia, and M. Antezza,
  Reconciliation of quantum local master equations with thermodynamics,
  New J. Phys. {\bf 20}, 113024 (2018).

\bibitem{Cattaneo-19}
  M. Cattaneo, G. Giorgi, S. Maniscalco, and R. Zambrini,
  Local versus global master equation with common and separate baths:
  Superiority of the global approach in partial secular approximation,
  New J. Phys. {\bf 21}, 113045 (2019).

\bibitem{DeChiara-20}
  A. Hewgill, G. De Chiara, and A. Imparato,
  Quantum thermodynamically consistent local master equations,
  Phys. Rev. Research {\bf 3}, 013165 (2021).

\bibitem{Farina-20}
  D. Farina, G. De Filippis, V. Cataudella, M. Polini, and V. Giovannetti,
  Going beyond local and global approaches for localized thermal dissipation,
  Phys. Rev. A {\bf 102}, 052208 (2020).

\bibitem{Datta}
  S. Datta,
  \textit{Electronic Transport in Mesoscopic Systems}
  (Cambridge University Press, Cambridge, 1995).

\bibitem{KaneMele}
  C. L. Kane and E. J. Mele,
  $Z_2$ topological order and the quantum spin Hall effect,
  Phys. Rev. Lett. {\bf 95}, 146802 (2005).

\bibitem{Rivas-14}
  {\'A}. Rivas, S. F. Huelga, and M. B. Plenio,
  Quantum non-Markovianity: Characterization, quantification and detection,
  Rep. Prog. Phys. {\bf 77}, 094001 (2014).

\bibitem{Breuer-16}
  H.-P. Breuer, E.-M. Laine, J. Piilo, and B. Vacchini,
  Colloquium: Non-Markovian dynamics in open quantum systems,
  Rev. Mod. Phys. {\bf 88}, 021002 (2016).

\bibitem{deVega}
  I. de Vega and D. Alonso,
  Dynamics of non-Markovian open quantum systems,
  Rev. Mod. Phys. {\bf 89}, 015001 (2017).

\bibitem{Abramowitz}
  M. Abramowitz, and I. Stegun,
  \textit{Handbook of mathematical functions with formulas, graphs, and mathematical tables}
  (Dover Publications, New York, 1964).

\end{thebibliography}
\end{document}